\documentclass[a4paper]{ar-1col}
\usepackage{epstopdf}
\usepackage[numbers]{natbib}
\usepackage{graphics}
\usepackage{numprint}
\def\gs{\mathrel{
   \rlap{\raise 0.511ex \hbox{$>$}}{\lower 0.511ex \hbox{$\sim$}}}}
\def\ls{\mathrel{
   \rlap{\raise 0.511ex \hbox{$<$}}{\lower 0.511ex \hbox{$\sim$}}}}

\jname{Annu. Rev. Part. Sci.}
\jvol{AA}
\jyear{YYYY}
\doi{10.1146/((please add article doi))}
\begin{document}

\markboth{Dolinski et al.}{Neutrinoless Double-Beta Decay}

\title{Neutrinoless Double-Beta Decay: Status and Prospects}

\author{Michelle~J.~Dolinski,$^1$ Alan~W.~P.~Poon,$^2$ and Werner~Rodejohann$^3$
\affil{$^1$Department of Physics, Drexel University, Philadelphia, Pennsylvania~19104, USA; email:  dolinski@drexel.edu}
\affil{$^2$Institute for Nuclear and Particle Astrophysics, Nuclear Science Division, Lawrence Berkeley National Laboratory, Berkeley, California~94720, USA; 
email: awpoon@lbl.gov}
\affil{$^3$Max-Planck-Institut f\"{u}r Kernphysik,
Postfach 10 39 80, 69029~Heidelberg, Germany; email: werner.rodejohann@mpi-hd.mpg.de}}

\begin{abstract}
Neutrinoless double-beta decay is a forbidden, lepton-number-violating
nuclear transition whose observation would have fundamental
implications for neutrino physics, theories beyond the Standard Model
and cosmology.  In this review, we summarize the theoretical progress
to understand this process, the expectations and implications under
various particle physics models, as well as the nuclear physics
challenges that affect the precise predictions of the decay half-life.
We will also provide a synopsis of the current and future large-scale
experiments that aim at discovering this process in physically
well-motivated half-life ranges.
\end{abstract}

\maketitle

\tableofcontents

\section{\label{sec:intro}Introduction}

The discoveries of neutrino oscillations, adiabatic lepton flavor
transformation and neutrino
mass~\cite{Fukuda:1998mi,Ahmad:2002jz,Eguchi:2002dm,Kajita:2016cak,McDonald:2016ixn}
furnished the first evidence of physics beyond the Standard Model
(SM).  Much is still to be learned about the neutrinos: the
mass-generation mechanism, absolute mass scale, $CP$-transformation
properties, and the question of whether they are Majorana
fermions~\cite{Majorana:1937vz}. The resolution of these unknowns
would extend our understanding of not only the underlying symmetries
that govern leptons, but also baryogenesis and the evolution of the
Universe.  If the SM-forbidden neutrinoless double-beta
($0\nu\beta\beta$) decay process were observed, it would directly
confirm lepton-number violation and the Majorana nature of
neutrinos~\cite{Schechter:1981bd}. Depending on the assumed mechanism,
vital information on the underlying model parameters could be
obtained. Interestingly, the range of possible models for
$0\nu\beta\beta$ decay extends from sub-eV neutrinos to multi-TeV
heavy particles, leading to a variety of potential consequences for
particle physics and cosmology. Without assuming any particular
driving decay mechanism, the searches for $0\nu\beta\beta$ decay are
searches for lepton-number violation whose observation would
demonstrate the breaking of a global conservation law of the SM.
These fundamental implications are the motivation for the prodigious
activities in the searches for experimental evidence and theoretical
underpinnings of this process.

Double-beta ($\beta\beta$) decay is an isobaric transition from a
parent nucleus ($A,Z$) to a daughter nucleus ($A,Z+2$) two nuclear
charges away.  In the two-neutrino double-beta ($2\nu\beta\beta$)
decay mode, two electrons and two electron-type antineutrinos
accompany the transition~\cite{GoeppertMayer:1935qp}:
\begin{equation}
(A,Z) \rightarrow (A,Z+2) + 2\,e^- + 2\,\bar{\nu}_e + Q_{\beta\beta},
\end{equation}
where $Q_{\beta\beta}$ is the energy released. It is a SM-allowed
second-order weak decay with a typical half-life of $>$10$^{19}$~y.
This decay mode was first deduced in a radiochemical experiment in
1950~\cite{Inghram:1950qv}, and subsequently observed in real time in
a dozen nuclei since the first laboratory measurement in the late
1980s~\cite{Elliott:1987kp,Moe:2014ioa}.  The readers are referred to
a previous article in the {\it Annual Review}
series~\cite{Saakyan:2013yna} for a comprehensive review.

No neutrinos are emitted in the SM-forbidden 
$0\nu\beta\beta$-decay mode:
\begin{equation}
\label{eq:nldbd}
(A,Z) \rightarrow (A,Z+2) + 2\,e^- + Q_{\beta\beta},
\end{equation}
in which the lepton number is violated by two units ($\Delta L = 2$).
The experimental search for this decay is extremely challenging, and
all previous attempts have returned empty-handed with the best current
half-life limits of $>$10$^{26}$~y.  The experimental difficulties are
matched by the theoretical ones; in particular, understanding the
nuclear physics aspects of the decay has been a persistent challenge.

There have been a number of review articles on $0\nu\beta\beta$ decay,
e.g.~Refs.~\cite{Moe:1994ss,Elliott:2002xe,Rodejohann:2011mu,DellOro:2016tmg}.
Our goal in this review is to capture some of the more recent
theoretical and experimental developments, as the current experiments
have reached a $0\nu\beta\beta$-decay half-life ($T_{1/2}^{0\nu}$)
limit in the range of $10^{25}$--$10^{26}$~y, and a worldwide program
to search for this decay with two orders of magnitude of improvement
in sensitivity is being pursued.

This article is organized as follows.  In Sec.~\ref{sec:particle}, the
particle physics motivations for the search for $0\nu\beta\beta$ decay
are provided.  The discussion will focus on the association of
$0\nu\beta\beta$ decay with lepton-number violation and neutrino mass,
as well as the mechanisms that could precipitate the decay.  The
interpretation of the observed signal would require knowledge of the
nuclear transition between the initial and final states.  The nuclear
matrix elements and other important aspects, such as quenching, are
the focus of Sec.~\ref{sec:nme}.  The design criteria that must be
considered in a $0\nu\beta\beta$-decay experiment are outlined in
Sec.~\ref{sec:ExpDesign}.  The broad range of detector technologies,
as well as the experimental status and prospects, are presented in
Sec.~\ref{sec:DetTech}. Section~\ref{sec:conclusions} is a summary.

\section{\label{sec:particle} Particle Physics Aspects}
\subsection{Why look for Lepton-Number Violation?}
As we stressed in the introduction, the searches for $0\nu\beta\beta$
decay are searches for lepton-number violation.  The lepton number is
an accidental global symmetry in the SM.  Theories beyond the SM
typically violate lepton number unless its conservation is forced by
the introduction of additional symmetries.

Lepton-number violation is most often introduced via a $\Delta L = 2$
Majorana mass term for standard or new neutrinos. Grand Unified
Theories (GUTs) normally require new neutral fermions; for instance,
the 16-dimensional spinorial representation of $SO(10)$ contains all
SM particles of a generation plus a right-handed neutrino. Those
particles are strongly motivated by the observation of neutrino
mass. Left-right symmetric theories or models that gauge the
difference of the baryon and lepton numbers $B-L$ also include
right-handed neutrinos with Majorana masses, at least in their minimal
formulations. Once the right-handed neutrinos are present, the gauge
symmetry of the SM necessarily implies the existence of light massive
Majorana neutrinos as a consequence of the seesaw mechanism.  An
example without any right-handed neutrinos is provided by
$R$-parity-violating supersymmetry, which contains $\Delta L = 1$
terms $\lambda' \, \ell Q D^c$ that couple the lepton doublet $\ell$,
the quark doublet $Q$, and the down-quark singlet $D$ superfields in
the Lagrangian.  The Majorana neutrino masses are generated at the
loop-level with two such vertices.

Hence, the arguments for lepton-number violation are strong and
plenty. Its strength, however, is model-dependent and needs to be
probed experimentally, just as the searches for baryon-number
violation in proton decay and neutron-antineutron oscillation
experiments.  A well-motivated framework often specifies the scales
that need to be tested, such as the non-zero minimal effective mass in
the inverted mass ordering of light neutrinos. A typical framework
with lepton-number violation would not only predict, within a more or
less definite range, the particular $0\nu\beta\beta$-decay half-lives
but also other observable quantities. Examples of such quantities
include the sum of the neutrino masses (as testable in cosmology)
within the standard light-neutrino paradigm, or the cross sections for
$eejj$ (same-sign di-electron plus di-jet) signals in the
heavy-particle exchange scenarios at the Large Hadron Collider
(LHC). These predictions allow for experimental checks and the
differentiation of individual mechanisms.

The simple observation that there is matter in the Universe implies
that some mechanism beyond the SM must exist to create
matter. Neutrinoless double-beta decay is obviously a process that
creates matter, and its observation is crucial for demonstrating
baryogenesis ideas.

It is interesting to consider the energy scales of the physics that
could be probed by lepton-number violation and proton decay. In the
standard light-neutrino mechanism, the $0\nu\beta\beta$-decay
half-life is proportional to $\Lambda^2$, where $\Lambda$ is the scale
of neutrino-mass generation, for example, the heavy-neutrino mass in
the seesaw mechanism. If heavy physics is responsible for the decay,
its half-life would be proportional to $\Lambda^{10}$ with $\Lambda$
being the mass of the heavy particles.  On the other hand, proton
decay has half-lives that are proportional to the GUT scale $\Lambda$
in the form of $\Lambda^{4}$ and $\Lambda^{5}$ for non-SUSY and SUSY
decay modes, respectively. These are obviously very different scales
to be tested, and it is difficult to generalize them in
model-independent statements.

\subsection{Neutrino Mass and Neutrinoless Double-Beta Decay}

The observation of neutrino oscillations demonstrated that neutrinos
have mass. The two main consequences from the impressive experimental
progress in the last two decades are: (i) the two different
mass-squared differences imply that all neutrino masses are different
with at least two of them being non-zero; and (ii) lepton mixing is
large.

The scalar and fermion content of the SM does not allow for neutrino
masses; hence, neutrino oscillations imply physics beyond the SM. It
is highly non-trivial to explain this ``new physics'' within a simple
paradigm.  In this ``3-Majorana neutrino paradigm,'' all phenomena
related to neutrino physics are generated by the neutrino mass matrix
$m_\nu = U\, {\rm diag} (m_1, \, m_2, m_3) \,U^T$,  
where $m_i$ are the real and positive neutrino masses and $U$ is the
Pontecorvo-Maki-Nakagawa-Sakata (PMNS) matrix containing three mixing
angles and three CP phases (one Dirac and two Majorana phases).

It is important to note that the smallest neutrino mass is not
currently known, and that two options for neutrino-mass ordering
exist: $m_3 > m_2 > m_1$ (normal ordering) and $m_2 > m_1 > m_3$
(inverted ordering). The cases in which the lightest neutrino mass is
much smaller than the heavier masses are denoted the normal or
inverted hierarchy.

There are altogether nine physical parameters in $m_\nu$, seven of
which appear in the effective mass\footnote{Only the least-known
  oscillation parameters $\theta_{23}$ and $\delta$ do not appear.}:
\begin{equation}\label{eq:meff}
\langle m_{\beta\beta} \rangle = \left| U_{ei}^2 \, m_i \right| . 
\end{equation}
All seven, except for the two Majorana phases $\alpha$ and $\beta$,
can be determined by other means---the absolute values of $U$ from
neutrino oscillations and the neutrino mass scale from direct
kinematic searches or cosmology.

In the light-neutrino exchange model, which has hitherto been the most
espoused in the physics community, the $0\nu\beta\beta$-decay
half-life is:
\begin{equation}\label{eq:T12}
  T^{0\nu}_{1/2} = \left( G \, |{\cal M}|^2 \, \langle m_{\beta\beta} \rangle^2
  \right)^{-1}
  \simeq 10^{27-28} \left( \frac{0.01 \, \rm eV}{\langle m_{\beta\beta} \rangle }
  \right)^2 \rm y\,.
\end{equation}
In this interpretation of $0\nu\beta\beta$ decay, it is a neutrino
mass experiment under the assumption that no other mechanism
contributes to lepton-number violation and that the neutrinos are
Majorana particles.

In Eq.~(\ref{eq:T12}), the phase-space factor $G \propto
Q_{\beta\beta}^5$ is of the order of
$10^{-25}$/(y~eV$^2$)~\cite{Kotila:2012zza,Stoica:2013lka}.  For the
nuclear matrix element (NME)---${\cal M}$---the approximation $|{\cal
  M}|^2 \sim 10$ is used.  From the experimental discussion in
Sec.~\ref{sec:DetTech}, it is clear that tonne-scale experiments are
needed to probe the physically interesting regime of
$T_{1/2}^{0\nu}\sim 10^{28}$~y and $\langle m_{\beta\beta}\rangle \sim
0.01$~eV.  The current limits on $\langle m_{\beta\beta}\rangle$ are
about 0.2~eV (see Tab.~\ref{tbl:curLimit}).

For the SM $V{-}A$ weak interaction, it is worth noting that any
observable connected to the Majorana nature of the neutrinos is
suppressed by the square of the neutrino mass divided by the energy
scale of the process~\cite{Balantekin:2018ukw}.  This is the reason
why $T_{1/2}^{0\nu}$ is so large compared to that in the
$2\nu\beta\beta$-decay process.

The neutrino mass can also be probed by direct kinematic searches,
such as the KATRIN~\cite{KATRIN} and ECHo~\cite{ECHo} experiments, as
well as by cosmological observations~\cite{Lattanzi:2017ubx}. The
kinematic searches and cosmological observations are sensitive to
\begin{equation}
m_\beta = \sqrt{|U_{ei}|^2  \, m_i^2}~\mbox{ and } ~\Sigma = m_1 + m_2 + m_3, 
\end{equation}
respectively. While the direct kinematic searches provide the most
model-independent approach to test the neutrino mass, they give the
weakest limits; the projected $m_\beta$ sensitivity in the KATRIN
experiment is 0.2~eV.

Cosmology gives the strongest mass limits in the sum of the neutrino
masses $\Sigma$.  But they depend on the data sets that need to be
combined in order to break the degeneracies of the many cosmological
parameters. The limits also become weaker when one departs from the
seven-parameter framework of $\Lambda$CDM plus neutrino mass (denoted
$\Lambda$CDM$+m_\nu$) to frameworks with more cosmological parameters.
The neutrino mass limits in exotic models of modified gravity are
difficult to quantify, but are expected to be weaker as well. The
current conservative limits on $\Sigma$ are about 0.3~eV.  The readers
are referred to the latest Planck data release~\cite{Aghanim:2018eyx}
for a detailed analysis of the cosmic microwave background and other
related data. It is noteworthy that a neutrino mass signal is quite
likely in future observations within the $\Lambda$CDM$+m_\nu$
framework, as well as in other moderate extensions; in particular, if
the Planck data are combined with the future Euclid and Square
Kilometre Array data~\cite{Sprenger:2018tdb,Brinckmann:2018owf}.  This
exciting prospect distinguishes the cosmological observations from
other approaches\footnote{The Project-8 experiment has the ambitious
  goal of probing $m_\beta$ to 40~meV, which would cover the
  inverted-ordering region~\cite{Esfahani:2017dmu}.}. Nevertheless,
let us repeat that the limits and constraints are hard to quantify in
exotic modifications of the minimal $\Lambda$CDM$+m_\nu$ model; the
combination of different data sets is prone to misinterpretations when
a multitude of systematic effects are present.

\begin{figure}[t]\vspace{-1.5cm}
\includegraphics[width=5.5in,height=4in]{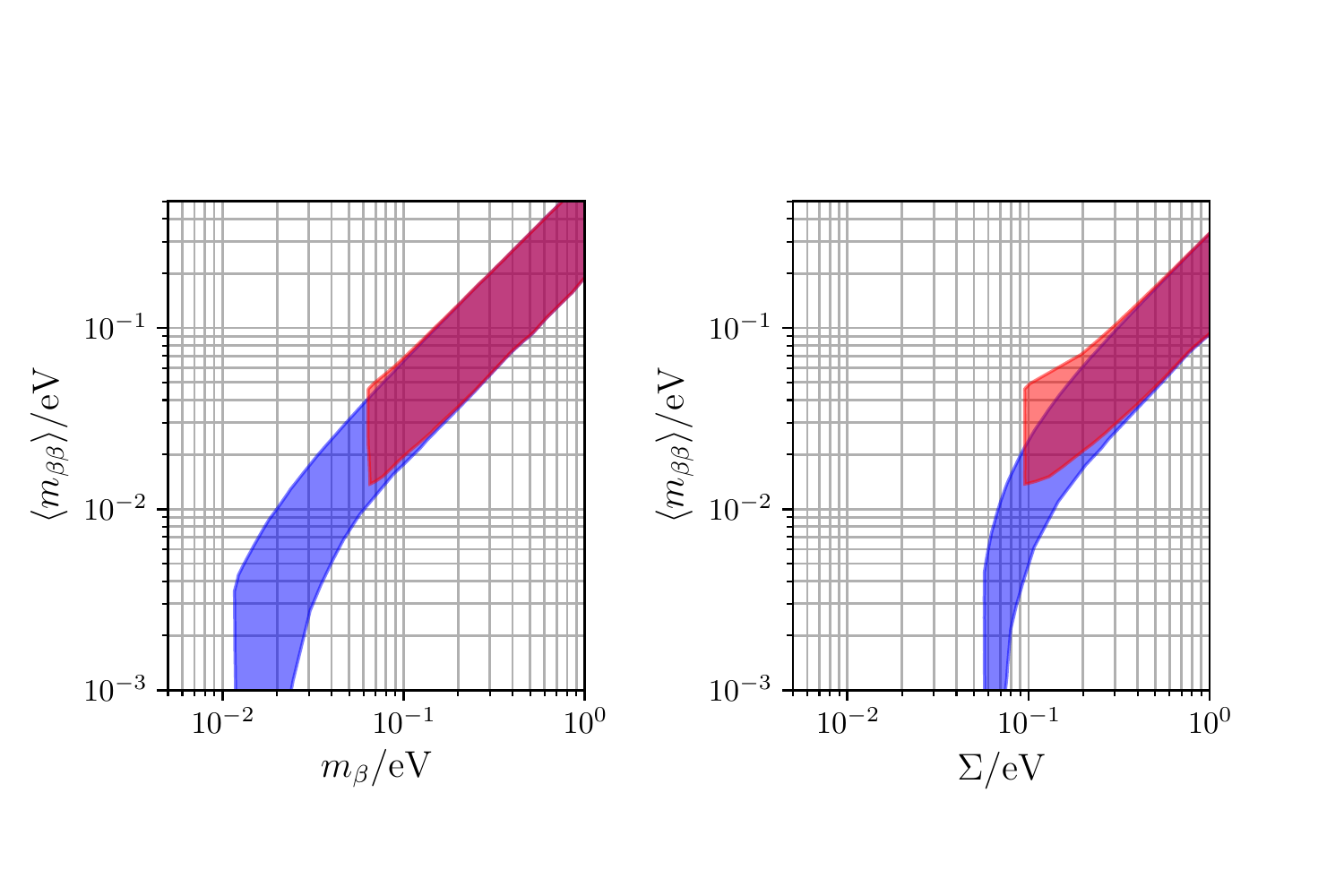}
\vspace{-1.5cm}
\caption{The effective mass $\langle m_{\beta\beta} \rangle$ versus
  the kinematic neutrino mass observable $m_\beta$, and the
  cosmological observable $\Sigma$.  The neutrino oscillation
  parameters are varied within their $3\sigma$ ranges.  The blue (red)
  area is for the normal (inverted) mass ordering.}
\label{fig:standard}
\end{figure}

The smallest neutrino mass and the Majorana phases are not
known. Varying them allows us to plot the three mass observables
against each other, which illustrates nicely the complementarity of
the different neutrino mass probes in Fig.~\ref{fig:standard}. In the
scenario of normal mass ordering with hierarchical masses, $\langle
m_{\beta\beta} \rangle$ is of the order of meV and can even vanish.
In the inverted-ordering scenario, there is a minimum value of about
0.013~eV~\cite{Pascoli:2002xq}. This value represents a physics goal
for the current and upcoming $0\nu\beta\beta$-decay experiments.

The current global fits of neutrino oscillation data favor the normal
mass ordering over the inverted one by more than
$3\sigma$~\cite{Esteban:2018azc}.  Small tensions in the values of the
oscillation parameters $\Delta m^2_{31}$ and $\theta_{13}$ obtained
from the long-baseline and the reactor experiments contribute to this
preference, as does an excess of upward-going $e$-like events in the
Super-Kamiokande atmospheric neutrino data. The current situation may
change; nevertheless, this preference has slowly strengthened with
time, as one would expect if it is indeed correct. It is important to
stress that normal ordering alone does not presuppose a tiny effective
mass; the smallest neutrino mass can still be sizable as normal {\it
  ordering} does not necessitate normal {\it hierarchy}.

Bayesian inference can be exploited to quantify the preference for
mass ordering by considering the cosmological and neutrino oscillation
constraints imposed on the available
data~\cite{Hannestad:2016fog,Schwetz:2017fey}.  The results depend
strongly on the choices of the prior (linear or logarithmic) and the
parameter space (neutrino masses, or the smallest mass and
mass-squared differences, or $\Sigma$ and mass-squared differences,
etc.). Normal ordering is most strongly preferred when the sampling is
performed for the three neutrino masses with logarithmic
priors~\cite{Gariazzo:2018pei}.

\begin{figure}[t]
\includegraphics[width=5in]{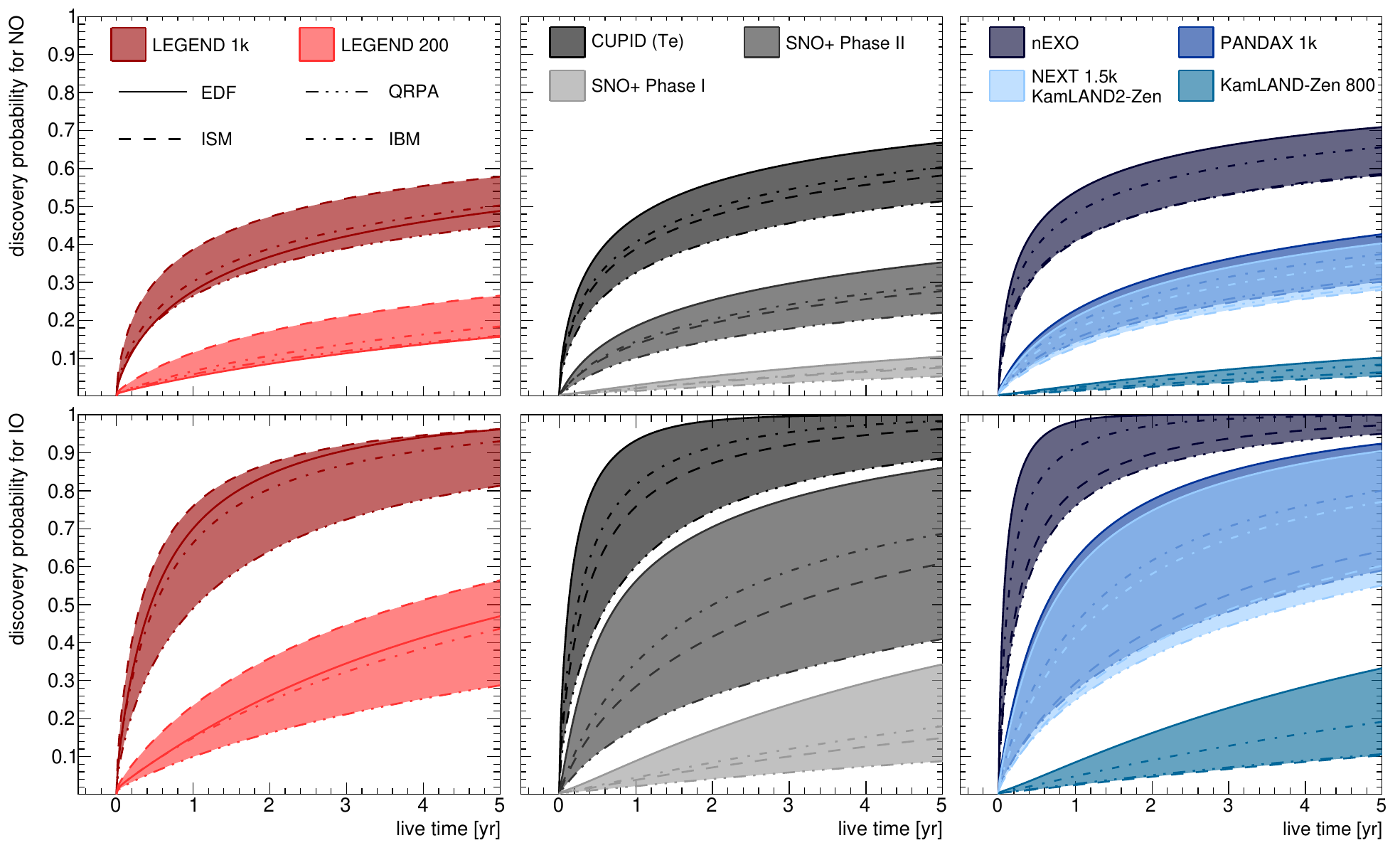}
\caption{Bayesian discovery probability for future experimental
  programs (CUPID, KamLAND-Zen, LEGEND, nEXO, NEXT, PANDA-X, SNO+) as
  function of running time. The upper (lower) plots are for the normal
  (inverted) ordering. Taken from \cite{Agostini:2017jim}.}
\label{fig:ABD}
\end{figure}

Insights can be gained by using oscillation and cosmology data to
obtain the probability distribution for $\langle m_{\beta\beta}
\rangle$, from which the discovery potential of future experiments can
be inferred~\cite{Caldwell:2017mqu,Agostini:2017jim,Ge:2017erv}.
Figure~\ref{fig:ABD} shows the ``Bayesian discovery probability,''
which corresponds to the chance of measuring a signal with a
significance greater than or equal to $3\sigma$.  The bands are due to
different assumptions in the nuclear matrix elements.  One can draw
optimistic conclusions from these Bayesian studies. There is a better
than 50\% discovery probability for normal ordering and almost unity
for inverted ordering for some of the future experiments.

Beyond SM neutrino physics, the presence of light sterile
neutrinos---prompted by LSND, MiniBooNE, short-baseline experiments,
and other anomalies---can change the picture dramatically.  The
additional contribution to the effective mass from sterile neutrinos,
$|U_{e4}|^2 \,m_4 \simeq |U_{e4}|^2 \,\sqrt{\Delta m_{41}^2}$, is of
the same order of magnitude as the minimal value of the effective mass
in the inverted ordering of active neutrinos. It shifts the half-life
distribution toward lower values~\cite{Ge:2017erv}; for normal
ordering, the shift is toward larger values.  The current situation on
light sterile neutrinos is confusing~\cite{Maltoni}; most likely, not
all hints are correct. Interestingly, the additional sterile-neutrino
parameters that enter the effective mass are the same ones that could
be responsible for the hints of active-to-sterile oscillation in
reactor antineutrinos, for which extensive experimental efforts are
being committed.  The readers are referred to
Ref.~\cite{Giunti:2019aiy} for a comprehensive review on sterile
neutrinos.

\subsection{\label{sec:alt}Alternative Mechanisms for Neutrinoless Double-Beta Decay}

In the light-neutrino mechanism discussed so far,
$0\nu\beta\beta$-decay searches are directly testing light physics.
Most alternative mechanisms are short-range mechanisms\footnote{See
  Ref.~\cite{Helo:2016vsi} for a recent discussion of alternative
  long-range mechanisms.}.  If the light neutrino were replaced by a
heavy neutrino (i.e.~with a mass $M_\nu$ larger than the
$0\nu\beta\beta$ scale of $|q| = 100$~MeV), the propagator would
become
\begin{equation}
\frac{M_\nu}{M_\nu^2 - q^2} \simeq \frac{1}{M_\nu} \ll \frac{1}{\sqrt{|q^2|}}\,.
\end{equation}

In fact, if only heavy particles mediate the decay, the amplitude of
the process would be
${\cal A}_{\rm heavy} \sim c/M^5 = (\tilde c/M)^5$,
where the mass scale $M^5$ is generally a combination of different
particle masses.  The corresponding amplitude for the standard one is
${\cal A}_{\rm st} \sim G_F^2 \, \langle m_{\beta\beta} \rangle/q^2$.
The current limit of $\langle m_{\beta\beta} \rangle$ yields
${\cal  A}_{\rm st} \sim (0.3 \, \rm TeV)^{-5}$, demonstrating that
lepton-number-violating TeV-scale physics would generate
$0\nu\beta\beta$-decay half-lives corresponding to the current
limits. This simple but illustrative and reasonably accurate estimate
is the basis of many works on testing alternative
$0\nu\beta\beta$-decay diagrams with the same-sign di-lepton
processes~\cite{Keung:1983uu} $pp \to ee jj$ at the LHC, or similar
processes at other colliders.

\begin{figure}[t]
\includegraphics[width=6in]{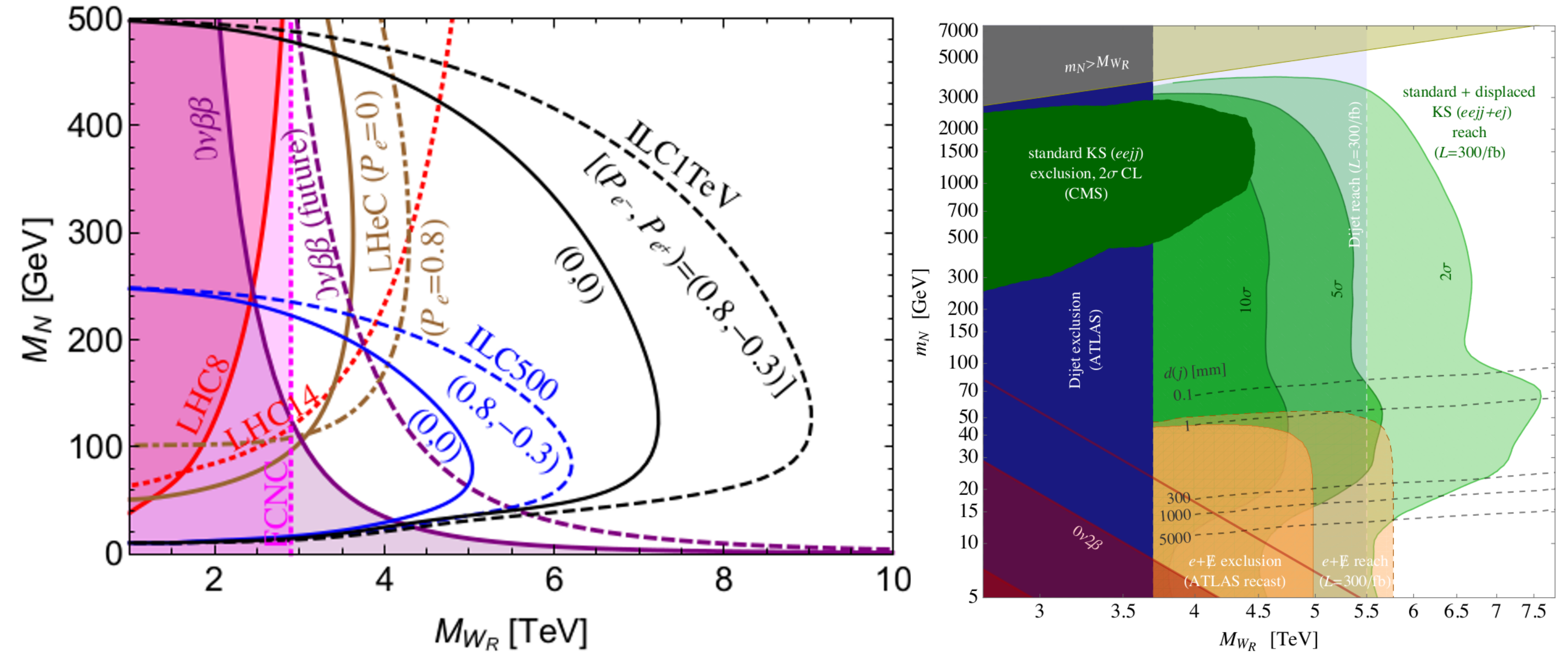} 
\caption{Several examples of constraining and testing alternative
  $0\nu\beta\beta$-decay mechanisms. Left: a comparison of various
  projected limits from future collider experiments on the
  right-handed neutrino and gauge boson mass; taken from
  Ref.~\cite{Biswal:2017nfl}. Right: the effects of displaced-vertex
  analysis on the same parameters; taken from
  Ref.~\cite{Nemevsek:2018bbt}. }
\label{fig:alt}
\end{figure}

Modifying Eq.~(\ref{eq:T12}), we can express the half-life for heavy
physics mechanisms very approximately (i.e.\ with much wider spread
than the light-neutrino expression) as
\begin{equation}
T^{\mathrm{heavy}}_{1/2} \sim 10^{27-28} \left( 
\frac{\tilde c/M}{\rm TeV}\right)^{10}
\rm y \,.
\end{equation}
A typical LHC test would work via the resonant production of a vector
boson and a Majorana fermion causes the subsequent lepton-number
violation.  The new particles in the $0\nu\beta\beta$-decay diagram
are not required to have the same or similar mass.  The fermion could
be much lighter than the vector boson, in which case the leptons and
jets would be of low energy and would escape detection in the
analysis. Displaced-vertex searches, as those shown in
Fig.~\ref{fig:alt}, are helpful in such
instances~\cite{Helo:2013esa,Antusch:2016ejd,Nemevsek:2018bbt}.
Future $e^+ e^-$ or $e p$ colliders have different characteristics in
particle kinematics, allowing experiments to probe different areas in
the parameter space.  Moreover, the polarization of the initial-state
fermions can help disentangle the chiral nature of the underlying
process~\cite{Lindner:2016lxq,Biswal:2017nfl}. Recent reviews on the
tests of neutrino mass models and lepton-number violation at colliders
can be found in Refs.\ \cite{Deppisch:2015qwa,Cai:2017mow}.

As an example, there are left-right symmetric theories that contain
heavy right-handed neutrinos $N_R$ and gauge bosons $W_R$ with mass
$M_{W_R}$.  Several diagrams for $0\nu\beta\beta$ decay arise in those
theories~\cite{Hirsch:1996qw,Barry:2013xxa,Cirigliano:2018yza}; for
instance, purely right-handed ones with $N_R$ and $W_R$ exchange, or
mixed diagrams with light-neutrino exchange in which one of the
currents is right-handed.  The electrons in these latter diagrams are
emitted with different helicities, which affect their angular
distribution. The energy distribution of the individual electrons is
also different, which in principle would allow the driving mechanisms
to be distinguished if the electrons can be tracked. Such an analysis
has been performed in the SuperNEMO project~\cite{Arnold:2010tu}.

A natural question to ask is whether $0\nu\beta\beta$ decay or
collider limits will provide better constraints on the relevant model
parameters. The answer depends on the various corrections that have
not been studied for all mechanisms.  As an example, a model with a
$Y$=1 $SU(2)_L$ doublet scalar and a singlet fermion was discussed in
Refs.\ \cite{Helo:2013ika,Peng:2015haa}. Including various
experimental and theoretical corrections, it was
shown~\cite{Peng:2015haa} that $0\nu\beta\beta$ decay would provide
better reach in the search for TeV-scale lepton-number-violating
interactions.  The size of many theoretical corrections is however not
completely understood and is a subject of debate. Nevertheless, LHC
and $0\nu\beta\beta$-decay searches are complementary approaches that
provide a consistency check in case of a discovery.

Lepton-number violation observed at TeV scale has interesting
cosmological implications. It is possible to translate an observed
cross section of a lepton-number-violating process at the LHC into
lepton-number-violating ``washout'' processes in the early
Universe~\cite{Frere:2008ct,Deppisch:2013jxa}. Any lepton asymmetry
generated by standard high-scale leptogenesis would typically be
washed out, making this baryogenesis mechanism ineffective. With
TeV-scale lepton-number violation, there is not really a need for
standard high-scale leptogenesis; nevertheless, the interesting
consequences of a TeV-scale observation of a $0\nu\beta\beta$-like
process are obvious.

The black-box, or Schechter-Valle theorem~\cite{Schechter:1981bd},
states that any diagram causing $0\nu\beta\beta$ decay will generate a
Majorana mass term for light neutrinos, which renders them Majorana
particles. However, this is generally a mass term generated by a
four-loop diagram that leads to a minuscule mass~\cite{Duerr:2011zd}
of the order of $\epsilon /(8\pi^2)^4 G_F^2 m_q^6/m_p \ls 10^{-29}$~eV, 
where we have used $m_q = 5$~MeV and $\epsilon \ls 10^{-7}$~(see
below).  In certain models, the connection of the $0\nu\beta\beta$
operator to a Majorana mass may be more direct.  Examples are diagrams
in which the particles that generate $0\nu\beta\beta$ decay are the
same ones that generate the neutrino mass in one-loop mechanisms.

One can write down a general Lagrangian responsible for
$0\nu\beta\beta$ decay via short-range mechanisms~\cite{Pas:2000vn}:
\begin{equation} \label{eq:Lshort}
{\cal L}_{\rm short} = \frac{G_F^2}{2 m_p} \left\{ \epsilon_1^X JJj + \epsilon_2^X J^{\mu\nu} J_{\mu \nu} j + \epsilon_3^X J^\mu J_\mu j + \epsilon_4^X J^\mu J_{\mu\nu} j^\nu + 
\epsilon_5^X J^\mu J j_\mu
\right\},
\end{equation}
where $J = \bar u (1\pm\gamma_5)d$, $j = \bar e (1 \pm \gamma_5) e^c$,
$J_{\mu\nu} = \bar u \frac i 2 [\gamma_\mu, \gamma_\nu] (1 \pm
\gamma_5)d$, $J_\mu = \bar u \gamma_\mu (1 \pm \gamma_5) d$, and
$j_\mu = \bar e \gamma_\mu (1 \pm \gamma_5) e^c$. The chirality of the
operators is encoded in $X = abc$, where $a,b,c$ is $L$ or $R$. If
some heavy physics generates lepton-number violation at a scale
$\Lambda$, one can generate effective operators as in
Eq.~(\ref{eq:Lshort}), where $\epsilon$ decreases with $\Lambda$.  The
product of three fermion currents illustrates that $0\nu\beta\beta$
decay in the short-range scenario can be described by dimension-9
operators.  While there are 24 independent operators
\cite{Graf:2018ozy}, only a few of them appear in non-exotic theories
beyond the SM.  The limits on the various $\epsilon^X$ are around
$10^{-7}$ to $10^{-10}$~\cite{Graf:2018ozy,Deppisch:2012nb}.  An
example within a well-known extension of the SM is a diagram with
heavy right-handed bosons $W_R$ and right-handed neutrinos mediating
the decay. In this case, $\epsilon_3^{RR} = V_{ei}^2 m_p
(m_{W}/M_{W_R})^4 /M_i$, where $V$ is the right-handed lepton mixing
matrix and $M_i$ is the masses of the heavy neutrinos.

Finally, we should mention that several mechanisms for
$0\nu\beta\beta$ decay may be present at the same time.  They could
even interfere with each other~\cite{Meroni:2012qf} as long as the
helicities of the emitted electrons allow for that.

\subsection{QCD Corrections}

QCD corrections to $0\nu\beta\beta$-decay diagrams are
important~\cite{Cirigliano:2018yza,Peng:2015haa,Gonzalez:2015ady,Mahajan:2013ixa,Arbelaez:2016uto,Gonzalez:2017mcg,Cirigliano:2017djv}. 
Naively, the effect is of order $\alpha_s/(16\pi^2) \log \Lambda^2/q^2 \simeq
0.1$, where $\Lambda$ is the scale of the mechanism and $q^2 \simeq
(100 \, \rm MeV)^2$ is the scale of the nuclear process. However, a
Fierz transformation might be needed to generate a color-singlet final
state that can be sandwiched between final-state nucleons. This
procedure generates operators with different Lorentz structures, which
can have drastically different nuclear matrix elements, and so
generates sizable corrections.  The standard light-neutrino exchange
diagram does not generate additional operators after applying the
Fierz transformation to the QCD-corrected one; thus, it is not
significantly affected by QCD corrections.  Applying QCD corrections
(as electroweak corrections are much
smaller)~\cite{Peng:2015haa,Mahajan:2013ixa,Gonzalez:2015ady}, the
operators in Eq.~(\ref{eq:Lshort}) are supposed to be run down to the
scale of $0\nu\beta\beta$ decay of about 100~MeV. Below 1~GeV, the
strong coupling becomes too large for applying perturbative
techniques.  There are ideas to cover this regime, but their numerical
impact is not clear yet.  QCD corrections to the long-range mechanisms
are expected to be smaller than the nuclear matrix element
uncertainties~\cite{Arbelaez:2016zlt}.

\subsection{Alternative Processes}

There have been suggestions of other decay modes to probe low-energy
lepton-number violation, to identify the neutrino mass nature, or to
entail both, over the years; Ref.~\cite{Balantekin:2018ukw} is a
recent summary.  The observation of these modes typically require
either non-relativistic
neutrinos~\cite{Long:2014zva,Berryman:2018qxn,Yoshimura:2011ri} or new
interactions~\cite{Rosen:1982pj,Rodejohann:2017vup}.  For heavier
neutrinos, the effects are observable only in certain mass ranges,
such as in meson or $W$ decays~\cite{Dib:2015oka}.

Neutrinoless double-electron capture \cite{Winter:1955zz}, $(A,Z) +2
e^-\to (A,Z-2)$ was of interest as an attractive alternative to
$0\nu\beta\beta$ decay, since there was the possibility of a resonant
enhancement if the initial and final-state energies are close to
degenerate \cite{Bernabeu:1983yb}. However, precise measurements of
the involved nuclear masses disfavor this
option~\cite{Eliseev:2012ih,Eliseev:2013ypa}. In addition, the decay
to excited states, neutrinoless double-positron decay, or various
combinations of electron capture and beta or positron decay suffer
from very low rates. Their observation would be a consistency check of
the underlying mechanism of lepton-number violation and could provide
useful information on the nuclear physics in neutrino-accompanied
processes.  Neutrinoless double-beta decay remains the most optimistic
channel to answer the pressing questions of lepton-number violation
and the neutrino nature.

\section{\label{sec:nme}Nuclear and Hadronic Physics Aspects }

The nuclear matrix element for $0\nu\beta\beta$ decay can be written
formally as
\begin{equation}
\langle {\rm final}| {\cal L}_{\ell-N}|{\rm initial} \rangle\,.
\end{equation}
What is needed for its evaluation are nuclear structure calculations
for the final and initial nuclear states, as well as a proper
transition from the fundamental lepton-quark Lagrangian to the
lepton-nucleon one ${\cal L}_{\ell-N}$. Both problems are essentially
independent from each other.  Determining the accuracy and
uncertainties of the various possible nuclear matrix elements can be
considered as the most challenging theoretical problem that hinders
precision studies of $0\nu\beta\beta$ decay in the event of a
discovery.

\subsection{\label{sec:hadr}Hadronization}
While the fundamental $0\nu\beta\beta$-decay Lagrangian is written at
the quark level, hadrons are present in the nucleus. Moreover,
operators need to be run from the fundamental high
lepton-number-violating scale down to the nuclear scale, and then
matched to the operators built from the hadronic degrees of freedom. A
problem is that the hadronic operators are often phenomenologically
written in terms of the form factors when the transition between
quarks and nucleons is made, as in
$\langle p | \bar u (1-\gamma_5)d|n\rangle = e^{-i(p - p')x} \bar u(p) 
(F_S (q^2) + F_{PS}(q^2) \gamma_5) u(p') \equiv J_{S-P}$. 
In this example, $u(p)$ and $u(p')$ are the spinors for the initial
and final-state neutron, and $p'- p = q$.  The $q^2$-dependence of
several form factors (particularly for scalar and pseudoscalar) is
unknown, as is their normalization (particularly for tensor). The
induced currents are also important, as one can see by considering the
following nucleon matrix element that is particularly relevant for
light-neutrino exchange:
\begin{equation}\label{eq:NuME} \nonumber
\begin{array}{c}
\langle p | \bar u \gamma^\mu (1-\gamma_5)d|n\rangle   \equiv J_{V-A}^\mu (x)
\\
 =  \bar u(p)  
\left(F_V (q^2) \gamma^\mu -i F_{W}(q^2)/(2m_p) \sigma^{\mu\nu} q_\nu - F_A \gamma^\mu \gamma_5 +  F_{P}(q^2)/(2m_p) \gamma_{5} q_\mu \right) u(p') e^{iqx}\,. 
\end{array}
\end{equation}
The normalization factors $F_i(q^2 = 0)$ are the coupling constants;
$F_V(q^2 = 0) = g_V$ and $F_A(q^2 = 0) = g_A$ are the vector and
axial-vector coupling constants.
 
One can use the language of chiral
symmetry~\cite{Cirigliano:2018yza,Peng:2015haa,Cirigliano:2017djv,Graesser:2016bpz,Cirigliano:2017ymo}
and effective field theory~\cite{Prezeau:2003xn} to identify the
necessary, i.e.\ the ones with the same symmetry structure under
chiral symmetry, and leading hadronic operators.  In chiral power
counting, the $\pi\pi \,ee$ operator is the leading one, corresponding
to a pseudoscalar interaction (two neutrons exchange a pion, which
converts from a $\pi^-$ to a $\pi^+$).  There is an ongoing effort
from the lattice QCD community to provide pion-level nuclear matrix
elements and the necessary low-energy coupling constants of the
operators~\cite{Nicholson:2016byl,Shanahan:2017bgi,Nicholson:2018mwc}.
In general, pion exchange implies a long-range interaction, which
overcomes the usual suppression of the short-range diagrams.  In
mechanisms that induce pseudoscalar operators at the tree level, such
as $R$-parity-violating SUSY, pion exchange can be expected to
dominate~\cite{Faessler:1996ph,Simkovic:1999re}.  A general effective
field theory framework that connects a chain of effective field
theories through various scales, including those at lepton-number
violation, electroweak-symmetry breaking, chiral-symmetry breaking and
$m_\pi$, has been formulated recently in
Ref.~\cite{Cirigliano:2018yza}. 

The induced pseudoscalar current that is proportional to $F_P$ in
Eq.~(\ref{eq:NuME}) is also connected to pion exchange.  It has been
argued~\cite{Simkovic:1999re} that the correction to the leading
Gamov-Teller matrix element is of the order $q^2/(q^2 + m_\pi^2) \sim
30\%$ in the light-neutrino case.  Recently, the short-range
contributions to the light-neutrino mechanism have been revisited
using the chiral language mentioned above in
Refs.~\cite{Cirigliano:2018hja,Wang:2018htk}.  Diagrams with
$\pi\pi\,ee$ couplings generate ultraviolet divergences in the $nn \to
pp \, ee$ amplitude, which can be cured by a counter term in the form
of a nucleon-nucleon contact term.  Recent lattice calculations also
identified a possibly important short-distance
contribution~\cite{Shanahan:2017bgi}. While leading in chiral power
counting, its size is currently not determined well and its impact is
not clear.

\subsection{General Aspects of the Nuclear Matrix Elements}

Focusing on the most-referenced light-neutrino mechanism, we consider the 
quark-level current $J^\mu = \bar u \gamma^\mu (1 - \gamma_5) d$.
As a second-order process, a time-ordered integration is needed:
\begin{equation}
\int d^4 x \, d^4y \, \langle f | T\{ J^\mu(x) J^\nu(y) \} | i \rangle 
\propto 
\sum\limits_n \frac{\langle f | J^\mu(\vec q)|n\rangle \, \langle n | J^\nu(-\vec q)|i\rangle }
{|\vec q| (E_n + |\vec q| + E_{e2} - E_i)} + (e2 \to e1, \mu \leftrightarrow \nu),
\end{equation}
which implies the introduction of a complete set of intermediate
states of energy $E_n$. All states up to about 100~MeV
contribute\footnote{In contrast, the SM-allowed $2\nu\beta\beta$ decay
  has only $1^+$ intermediate states (as two real neutrinos are
  emitted) with energies up to $Q_{\beta\beta}$ of a few MeV.}.  The
impulse approximation for the nuclear current $J^\mu_{V-A}$ in
Eq.~(\ref{eq:NuME}) sums over the individual free-nucleon matrix
elements, i.e.~only one nucleon experiences the weak decay without
interference from the surrounding nuclear medium. The form factors
(see Sec.~\ref{sec:hadr}) need to be properly expanded in a
non-relativistic form. Various other approximations would then lead to
the general formula in Eq.~(\ref{eq:T12}).

The nuclear matrix element (NME) is 
\begin{equation}
{\cal M} = {\cal M}_{GT} - \frac{g_V^2}{g_A^2}{\cal M}_{F} + {\cal M}_T ,
\end{equation}
where the Fermi matrix element ${\cal M}_{F}$ depends on the integral
over $|\vec q|$ of $F_V(\vec q \,^2)$ in its non-relativistic
approximation, whereas the Gamov-Teller matrix element ${\cal M}_{GT}$
depends on the corresponding integrals over linear combinations of
$F_{A,P,W}(\vec q \,^2)$; see Refs.~\cite{Engel:2016xgb,Graf:2018ozy}
for the explicit expressions.  The tensor matrix element ${\cal
  M}_{T}$ can be neglected.  As an example, the Gamov-Teller matrix
element, which is the leading one, can be written as
\begin{equation}
{\cal M}_{GT} = g_A^2 \frac{2 R}{\pi } \int\limits_0^\infty d|\vec q|\, 
|\vec q| \langle f | \sum\limits_{a,b} \frac{j_0(|\vec q|r_{ab}) \, h_{GT}(|\vec{q}| 
\vec \sigma_a \cdot \vec \sigma_b)}{|\vec q| + \bar E - (E_i + E_f)/2} \tau^+_a \tau^+_b 
|i \rangle , 
\end{equation}
where $R$ is the nuclear radius of $1.2 A^3$ fm, $j_0$ is the Bessel
function, $h_{GT}$ is a combination of $F_{A,P,W}$ properly expanded,
and $\vec r_{ab}$ is the distance between the two decaying nucleons.
Short-range correlations may be important, particularly for
short-range mechanisms. The repulsion at short distances can be
phenomenologically described by the UCOM, Jastrow, Argonne or Bonn
potentials, with which the operators in the nuclear matrix elements
are multiplied.

\begin{figure}[t]
  \includegraphics[width=3in]{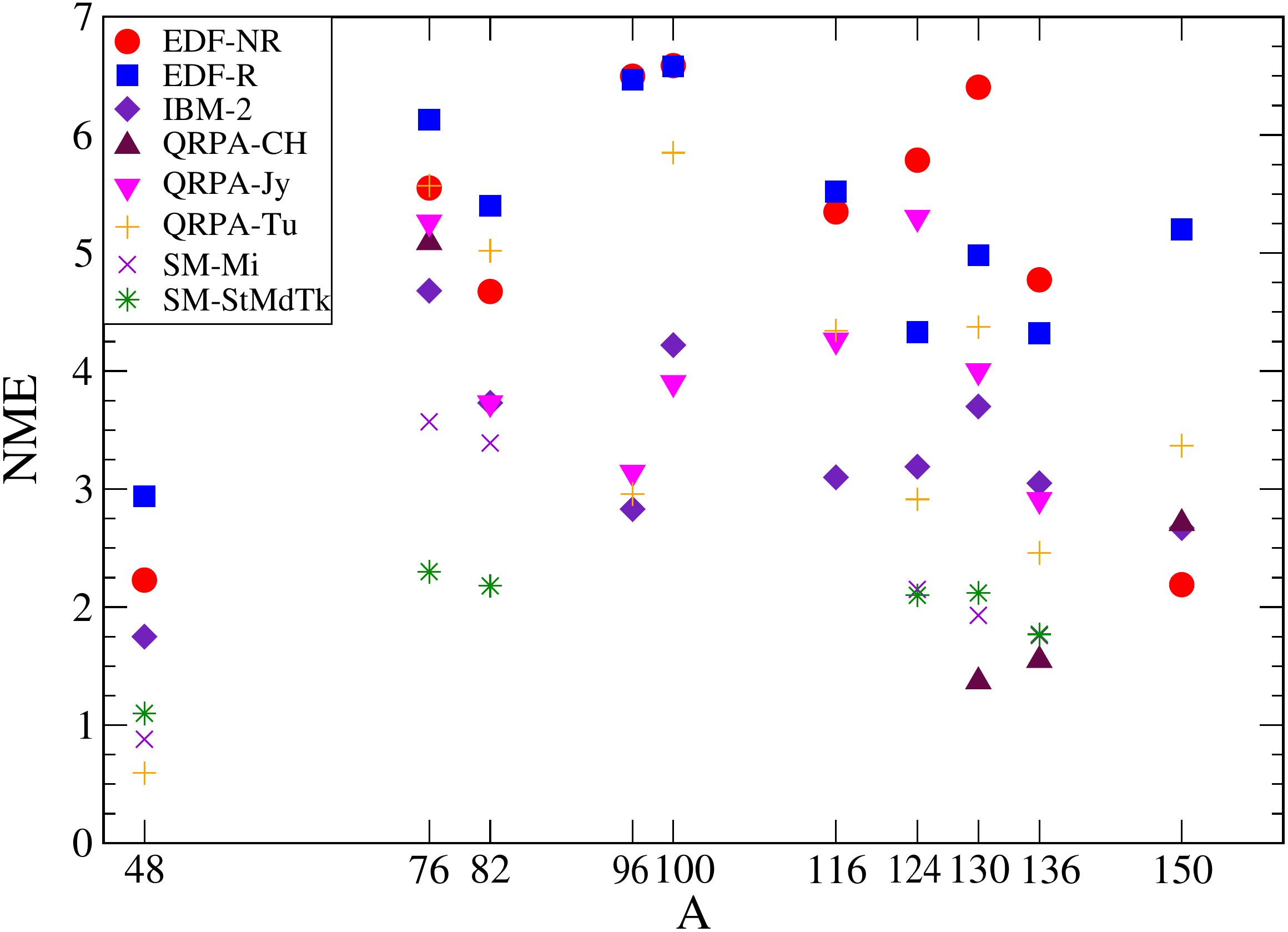}
  \caption{A representative compilation of nuclear matrix element
  calculations with an unquenched $g_A = 1.27$ for different
  isotopes. See Ref.\ \cite{Engel:2016xgb} for details and
  references.}
\label{fig:Engel}
\end{figure}

The difficulty of NME calculations is to know the initial and
final-state nuclear wave functions, a many-body problem that has no
exact solution.  Several approaches to the problem exist, and are
summarized in recent
reviews~\cite{Engel:2016xgb,Vergados:2012xy,Vergados:2016hso}.  The
status of NME calculations for the different approaches is depicted in
Fig.~\ref{fig:Engel}.  We summarize the main approaches in the
following.

The energy-density functional (EDF) and the generator coordinate
methods (GCM) mix many mean fields with different
properties~\cite{Rodriguez:2010mn,Vaquero:2014dna,Yao:2014uta},
whereas the other methods use simple mean fields that the states and
orbitals feel. Minimization of the energy functional finds the ground
states.  A large number of single-particle states are included and
their collective motion is treated, but only a few selected
correlations are used, possibly leading to an overestimation of the
NME. The Projected Hartree-Fock-Bogoliubov Method (PHFB) is a related
approach~\cite{Rath:2010zz}.

The nuclear shell model (NSM) does not use the full Hilbert space of
the nucleon states, but only those in a ``valence space'' near the
Fermi surface~\cite{Menendez:2008jp,Horoi:2015tkc}. The limited number
of active nucleons and oscillator shells means that the low-lying
states can be well-described and reproduced, although the effects of
pairing correlations may not be fully captured and may lead to an
underestimation of the NME. Indeed, enlarging the configuration space
would increase the matrix elements~\cite{Iwata:2016cxn}.

The interacting boson model (IBM) features nucleon pairs represented
as bosons with certain quantum numbers and features a truncation of
the full shell-model space to a subspace.  More shells are used than
in the NSM, but with fewer
correlations~\cite{Barea:2009zza,Barea:2013bz,Barea:2015kwa}.  The
description is typically more phenomenological than the other methods,
and relies more on adjusting the model parameters to match the
observables.

The quasi-particle random-phase approximation (QRPA) contains few
correlations but a large number of single-particle
orbits~\cite{Fang:2015zha,Mustonen:2013zu,Jokiniemi:2018sxd}. The
proton-neutron interaction quantified by a parameter $g_{pp}$ should
equal $1$ in an exact calculation and diagonalization; it is fixed to
a value that reproduces the measured $2\nu\beta\beta$-decay
half-lives. Also the particle-hole coupling parameter can be fixed by
observables.

The ``ab initio'' methods are a recent and promising line of
development. All nucleons are taken as degrees of freedom and
interactions are fitted from the data involving nucleons or small
nucleon systems.  These approaches are currently limited by the
availability of computing power. However, the recent results within
the NSM for light isotopes~\cite{Pastore:2017ofx} and the
$0\nu\beta\beta$-decay candidate $^{48}$Ca are
encouraging~\cite{Yao:2018qjv}.

All approaches miss certain features.  Naively one expects the lack of
configurations underestimates the NME, while the lack of correlations
overestimates them, which is what the distribution in
Fig.~\ref{fig:Engel} seems to confirm. There is hope that the
calculations will converge as their respective shortcomings are
overcome by future improvements. We refer the readers to the
authoritative nuclear physics review in Ref.~\cite{Engel:2016xgb},
which discusses those attempts in detail.

The uncertainties in the nuclear matrix elements are difficult to
quantify. Some effects would shift all matrix elements, for example,
the possible quenching of $g_A$ (see Sec.~\ref{sec:ga}); while others
are only applicable to certain models, such as the particle-particle
coupling within QRPA.  Although it is possible to study the effects by
varying the nuclear model parameters~\cite{Faessler:2008xj}, it is
less clear how to quantify the shortcomings of the models in a
systematic way. These desperately needed studies are underway as the
associated uncertainties are expected to be larger than those from
varying the model parameters. A multi-isotope $0\nu\beta\beta$-decay
program would surely help quantify and understand the current
discrepancies.

\subsection{\label{sec:ga}Quenching}

With the Gamov-Teller matrix element ${\cal M}_{GT}$ the leading one
in the light-neutrino exchange case, the nuclear matrix element is to
a good approximation proportional to $g_A^2$, and $T_{1/2}^{0\nu}$ is
proportional to $g_A^{-4}$.  Quenching denotes the reduction of $g_A$
that is necessary to reproduce the observable quantities of nuclear
decays~\cite{Deppisch:2016rox}, particularly $\beta$ and
$2\nu\beta\beta$ decays; see Ref.~\cite{Suhonen:2017krv} for a
review. In addition, low-energy forward-angle charge-exchange reaction
tests of the Ikeda sum rule confirm a reduced Gamov-Teller
strength~\cite{Frekers:2013yea}, as do the spectral measurements of
forbidden $\beta$ decays~\cite{Bodenstein-Dresler:2018dwb}.  A reduced
$g_A$ implies a longer $T_{1/2}^{0\nu}$, which is undesirable for
experimental
searches~\cite{Barea:2013bz,Barea:2015kwa,DellOro:2014ysa}. Other
alternative $0\nu\beta\beta$-decay mechanisms would also be affected
by quenching, though possibly to a lesser extent as ${\cal M}_{GT}$
may not be the leading element.

It is important to stress that there is no ``theory'' of
quenching. The value of $g_A$ is used as an adjustment to bring
observations in agreement with calculations. When the strength of the
Gamov-Teller operator needs to be reduced, one reduces the axial
coupling constant from its free nucleon value of $1.27$.  Possible
origins of quenching are nuclear-medium effects, many-body currents,
or the inherent shortcomings of the nuclear many-body models. A
possibly important observation is that $\beta$ and $2\nu\beta\beta$
decays have energy scales of order MeV, i.e.\ much smaller than the
$0\nu\beta\beta$ scale of order 100 MeV.  Low-energy processes may
require more quenching as the missing particle-hole excitations in the
models may shift the Gamow-Teller strength to higher energies. Thus,
less or no quenching might be needed in $0\nu\beta\beta$
decay. However, conflicting statements in the literature exist. The
dependence of quenching on the nuclear calculations can be
demonstrated by analyzing the $2\nu\beta\beta$ electron-energy
spectra~\cite{KamLAND-Zen:2019imh}, which allows the extraction of the
sub-leading higher-order contributions to the matrix
elements~\cite{Simkovic:2018rdz}.

Non-nucleon degrees of freedom or many-nucleon currents may also shed
light on the
issue~\cite{Menendez:2011qq,Engel:2014pha,Ekstrom:2014iya}.  In
$\beta$ and $2\nu\beta\beta$ decays, long-range pion exchange reduces
the matrix elements significantly, whereas a reduction of only 10-30\%
was observed in $0\nu\beta\beta$ decay (as pion exchange contributes
less at higher momenta). There are also indications that muon capture
on nuclei requires less quenching, which again implies an energy
dependence of the effect.

To sum up, recent studies indicate that there is less quenching
necessary (not more than 20-30\%) in processes with large momentum
transfer such as $0\nu\beta\beta$ decays.  This reduction would
correspond to an increase of $T_{1/2}^{0\nu}$ by a factor of about
$2$-$3$. However, there is not yet consensus in the literature on this
issue, and further experimental inputs and improvements in the
calculations are desperately needed.

\subsection{Experimental Tests of the Nuclear Matrix Elements}

Hadronic charge-exchange reactions, whose transition matrix
elements\footnote{While the charge-exchange reactions are mediated by
  nuclear 
  to the same spin, isospin and multipole operators. Thus, they can be
  used as tests of the weak interaction; in particular, as tests of
  the isospin response. } (related to the products of two beta-decay
Gamow-Teller matrix elements) can be accessed through reactions in the
$\beta^-$ and $\beta^+$ directions~\cite{Frekers:2018edj}, provide a
good test for the matrix elements in $2\nu\beta\beta$ decay. The
neutrinoless mode and its NME problem can benefit from such nuclear
structure measurements. For instance, the determination of the neutron
occupancies in $(p, t)$ two-nucleon transfer
experiments~\cite{Freeman:2007mm} has significantly influenced the
QRPA calculations of the NMEs~\cite{Menendez:2009xa}.  One is normally
interested in the Gamow-Teller operators, and has to choose reactions
with small momentum transfer at the forward angles.  By properly
choosing the kinematics, charge-exchange reactions can also probe the
transition strengths to the intermediate states beyond $1^+$, although
no relative phase information can be accessed.

The NUMEN collaboration provides a new approach
recently~\cite{Cappuzzello:2018wek}.  The goal is to use
heavy-ion-induced double charge-exchange reactions to test the
second-order isospin response. Even at the forward angles, sizable
momenta are transferred. Other similarities to $0\nu\beta\beta$ decay
include complex nuclear medium effects and off-shell intermediate
states of the reaction\footnote{Measurements of the double
  Gamow-Teller giant resonance are suggested to have a linear
  correlation to the $0\nu\beta\beta$-decay matrix
  element~\cite{Shimizu:2017qcy}; see however
  Ref.~\cite{Simkovic:2018hiq}.}.  The first measurements of the
$^{40}$Ca($^{18}$O,$^{18}$Ne)$^{40}$Ar reaction were
performed~\cite{Cappuzzello:2015ixp} to demonstrate the experimental
principle. Work is underway to probe reactions that involve the
isotopes in $0\nu\beta\beta$-decay searches. Apart from various
nuclear structure information, the quenching issue can also be
addressed. The latter is also possible in muon-capture
reactions~\cite{Hashim:2017qcw}.

\section{\label{sec:ExpDesign}Experimental Design Criteria}

The observables in direct searches of $0\nu\beta\beta$ decay are the
kinematic parameters of the two emitted electrons.  A typical
experiment measures the total energy ($E$) of the two electrons, and
may have the capability of reconstructing the individual electron
paths (tracking) to reject backgrounds based on event topology.  The
observed $0\nu\beta\beta$-decay signal is a monoenergetic peak at
$Q_{\beta\beta}$ as there are no antineutrinos emitted in the decay.
Since $Q_{\beta\beta}$ is well-measured, usually in high precision
atomic traps, the signal search can be performed over a narrow energy
window around $Q_{\beta\beta}$; the width of this ``region of
interest'' (ROI) is selected based on the energy resolution of the
detector.  The number of candidate events $N$ observed in the ROI is:
\begin{equation}
	N = \ln(2) \frac{N_\mathrm{A}}{W}\left(\frac{a \,  \varepsilon \, M \, t }{T_{1/2}^{0\nu}} \right), 
\end{equation}
where $N_\mathrm{A}$ is the Avogadro's number, $W$ is molar mass of
the source, $a$ is the isotopic abundance of the parent isotope,
$\varepsilon$ is the detection efficiency of the signal in the ROI,
and $t$ is the measurement time.  The last factor of this expression
captures the choices that an experimenter can make in designing an
experiment.

The sensitivity to the half-life obviously would depend on the total
number of counts in the ROI, some of which may be background events:
\begin{equation} 
\label{eq:Tsens}
(T_{1/2}^{0\nu}) \propto \left\{ 
\begin{array}{ll} 
 \displaystyle a \, M \, \varepsilon \, t &  \mbox{background free,} \\ 
 \displaystyle a \, \varepsilon \, \sqrt{\frac{M \, t}{B \, \Delta E}} &  \mbox{with background,} 
\end{array} 
\right. 
\end{equation}
where $\Delta E$ is the detector energy resolution, and $B$ is the
background index, normalized to the width of the ROI, source mass, and
measurement time, e.g.\ in units of (keV~kg~y)$^{-1}$.  This
expression shows clearly the advantage of a background-free
experiment, as the $T_{1/2}^{0\nu}$ sensitivity would scale linearly
with $t$ as opposed to $\sqrt{t}$ in the presence of backgrounds.  In
this section, we will discuss some of the design considerations in a
$0\nu\beta\beta$-decay experiment, including the choice of isotopes,
sources of backgrounds, as well as their mitigation and elimination.

\subsection{Isotope Choices}

\begin{table}[t]
\label{tbl:isotopes}
\begin{center}
\caption{Characteristics of commonly used $\beta\beta$-decay isotopes.  The isotopic abundances are obtained from Ref.~\cite{IUPAC:2011}.}
\begin{tabular}{r c c} \\ \hline\hline
Isotope	&  Natural abundance (\%)	 &  $Q_{\beta\beta}$ (MeV)\\ \hline
$^{48}$Ca		&  0.187	& 4.263 \\	
$^{76}$Ge	& 7.8		& 2.039 \\	
$^{82}$Se		& 8.7		& 2.998 \\	
$^{96}$Zr		& 2.8		& 3.348 \\	
$^{100}$Mo	& 9.8		& 3.035 \\	
$^{116}$Cd	& 7.5		& 2.813 \\	
$^{130}$Te	& 34.08	& 2.527 \\	
$^{136}$Xe	& 8.9		& 2.459 \\	
$^{150}$Nd	& 5.6		& 3.371 \\ \hline	
\end{tabular}
\end{center}
\end{table}

There are 35 isotopes capable of $\beta\beta$
decay~\cite{Tretyak:2002dx}, but not all of them are suitable as a
candidate isotope for direct searches of $0\nu\beta\beta$ decays.
Table~\ref{tbl:isotopes} lists the characteristics of some of the
isotopes that have been deployed in experiments.  Given
Eq.~(\ref{eq:Tsens}), an ideal isotope should have a high isotopic
abundance (large $a$), can be deployed in large quantity (large $M$)
as high resolution detectors (small $\Delta E$) under low-background
conditions (small $B$).  Unfortunately, such an isotope does not exist
and experimenters have to make design choices to optimize a subset of
these parameters.

The most critical consideration is the potential sources of
backgrounds.  An irreducible background to $0\nu\beta\beta$-decay
search is the $2\nu\beta\beta$-decay electrons; they are
indistinguishable from those in the $0\nu\beta\beta$-decay mode in the
ROI.  One way to mitigate this background is to deploy an isotope 
with a long $2\nu\beta\beta$-decay half-life.  The ratio of the
$0\nu\beta\beta$-decay signal to the $2\nu\beta\beta$-decay background,
$S/B$, is approximately~\cite{Elliott:2002xe}:
\begin{equation}
\frac{S}{B} \propto \left( \frac{Q_{\beta\beta}}{\Delta E}\right)^6 \,  \frac{T_{1/2}^{2\nu}}{T_{1/2}^{0\nu}},
\end{equation}
which indicates the importance of an excellent detector energy
resolution for isotopes that have shorter $2\nu\beta\beta$-decay
half-lives.

Primordial radioisotopes from the U and Th chains are ubiquitous in
the detector construction materials.  The most troublesome one is
$^{208}$Tl.  Its 2615-keV~$\gamma$-ray line lies above
$Q_{\beta\beta}$ for a number of $\beta\beta$-decay isotopes, and can
deposit energy extraneously within the ROI.  Another problematic
background comes from $^{222}$Rn, whose progeny $^{214}$Bi emits a
$\beta$~electron with an energy up to 3270~keV.  An ideal
$0\nu\beta\beta$-decay isotope candidate would have a $Q_{\beta\beta}$
high enough to avoid these backgrounds.

The detection efficiency of the $0\nu\beta\beta$-decay signal can be
significantly enhanced if the source material is integrated as the
detector medium.  As the path lengths of the two signal electrons are
much shorter than the size of the active medium in such a coalesced
configuration, calorimetry with excellent energy resolution is
possible.  When the source material is external to the detector, the
probability of at least one of the two electrons escaping detection or
with degraded energy increases due to self-absorption.  The main
advantage of this external-source configuration is the possibility of
superior tracking and effective background rejection, but at the
expense of energy resolution.

To reduce the cost of an experiment, an ideal source material should
be readily available in its natural form and the candidate isotope
within it should have a high natural abundance.  The cost of isotope
enrichment typically depends on the isotopic abundance of the starting
material---the higher the natural abundance, the lower the cost.  If
the natural abundance is high enough, isotope enrichment may be
unnecessary as has been demonstrated in the case of
$^{130}$Te~\cite{Alduino:2017ehq}. Reference~\cite{Saakyan:2013yna}
provides a succinct summary of the enrichment of $\beta\beta$-decay
isotopes.

\subsection{\label{sec:expbgd}Backgrounds}

The $T_{1/2}^{0\nu}$ discovery potential would shrink substantially in
the scenario of a non-vanishing background index.  For the next
generation of experiments to reach a discovery potential of
$T_{1/2}^{0\nu}\sim10^{28}$~y, an extremely stringent background index
of ${<}0.1$~count/(FWHM~t~y), where FWHM is the full width at half
maximum of the detector resolution at $Q_{\beta\beta}$, is necessary.
The readers are referred to a comprehensive review of backgrounds in
sensitive underground experiments in
Refs.~\cite{Heusser:1995wd,Formaggio:2004ge}.  The following is a
brief introduction.

As we have discussed above, a careful choice of target isotope and
detector technology could diminish the impact of the irreducible
$2\nu\beta\beta$-decay background on the discovery potential.
Similarly for the omnipresent solar neutrinos, their impact can be
mitigated by a high mass loading of the decaying isotope in the target
medium to improve the ratio of the signal to the neutrino-electron
elastic scattering background.  This is particularly important for
large (kilotonne scale) liquid scintillator detectors.

In $0\nu\beta\beta$-decay experiments, there are several types of
backgrounds that can be controlled through careful design and vigilant
implementation.  Trace amount of radioisotopes from the natural U and
Th chains must be kept to a minimum in any materials close to active
detector volume.  Other pervasive natural radioactivities, such as
$^3$H, $^{14}$C and $^{40}$K, have lower decay energies and do not
impinge on $0\nu\beta\beta$-decay searches. The techniques to produce
radiopure materials for mechanical support are constantly being
explored and refined; for example, electroformed copper and
alloys~\cite{Hoppe:2014nva,Suriano:2018nrb}, and
polymers~\cite{Majorovits:2017cqj}.  Radioassay results from prior
generations of low-background
experiments~\cite{Leonard:2007uv,Leonard:2017okt,Abgrall:2016cct} are
now readily accessible as online databases~\cite{Loach:2016fsk} to aid
the material selection process for future experiments.  Even when
intrinsically radiopure construction materials have been identified,
extreme care to maintain their cleanliness is essential.  For example,
exposure to $^{222}$Rn would result in increased $\alpha$ and $\beta$
emitter backgrounds on the surface or in the bulk of the unprotected
components.

Natural radioactivities far away from the active detector volume,
including $\gamma$~rays from the primordial chains and neutrons from
($\alpha,n$) reactions originated from the rock wall of the
underground laboratory, can be blocked by passive shielding with clean
lead or copper, water or liquid cryogen.  The latter two options may
also allow the shielding medium to serve as an active veto to reject
cosmic rays.

Cosmic-ray muons ($\mu$) can induce several types of backgrounds in a
$0\nu\beta\beta$-decay experiment.  For experiments at deep
underground laboratories, prompt muon interactions in the detectors do
not usually pose any background concerns. These interactions typically
deposit a large amount of energy and can be vetoed easily.  The
activation of long-lived isotopes and the production of secondary
neutrons are the main worries.  Muons can induce these backgrounds via
different mechanisms: $\mu^-$~capture in
nuclei~\cite{Suzuki:1987jf,Macdonald:1965zz,HEISINGER2002357};
$\mu$-nucleon quasi-elastic scattering; electromagnetic showers; and
photo-neutron production through virtual photon exchange. High-energy
neutrons produced in inelastic neutron scattering $(n,n^\prime\gamma$)
are also a source of background in $0\nu\beta\beta$-decay experiments~\cite{Boswell:2012dm,Negret:2013fsa}.

Numerous theoretical and experimental studies have been performed to
determine the production yield of these radioisotopes in materials
commonly used in dark matter and $\beta\beta$-decay searches have
been exposed to cosmic rays at or above the Earth's surface; see e.g.
Refs.~\cite{Wang:2015pxa,Baudis:2015kqa,Armengaud:2016aoz,Ma:2018tdv}
on target materials, and
Refs.~\cite{HEISINGER2002345,Cebrian:2017oft,Zhang:2016rlz} on
construction materials.  There are two strategies to mitigate these
activated backgrounds: to minimize the exposure to cosmic rays on surface
and to let the materials ``cool down'' underground after such
exposure.  However, it would be impractical to wait for certain
long-lived radioisotopes to decay to an acceptable activity.

The backgrounds from cosmogenic production of radioisotopes in situ
during the experiment are difficult to identify as their decays could
occur long after the initial muons.  Although this is an irreducible
background, its impacts can be mitigated by simply deploying the
experiment at a greater depth.  This type of backgrounds is of
particular concern to experiments in which the $\beta\beta$-decay
isotopes are dissolved in a large volume of host medium (e.g.~liquid
scintillator~\cite{Abe:2009aa}) given the large mass and the broad
energy spectrum of the activated products.  Experiments with tracking
or event position reconstruction capabilities can reject these
backgrounds by temporal and spatial correlations; see
e.g.\ Ref.~\cite{EXO200::2015wtc}.

\subsection{Detection Strategies}

As with any search for new physics, the primary goal of the detector
design is to discriminate between signal and backgrounds effectively
while maintaining high signal detection efficiency. The most common
way to achieve this discrimination is via energy resolution, which is
generally intrinsic to the detection medium. For the next generation
of experiments, it will also be essential to maximize the discovery
potential. This means actively showing that any observed signal is not
only consistent with the expected $0\nu\beta\beta$-decay signal but
also inconsistent with the measured backgrounds. The pitfall of
relying on energy alone to make a discovery claim is illustrated by
Ref.~\cite{KlapdorKleingrothaus:2004wj}.

Neutrinoless double-beta decay has a characteristic event topology
with the emission of two ${\sim}$MeV electrons. Low-density-gas
tracking detectors can in principle resolve the two electron tracks,
leaving only the irreducible background from $2\nu\beta\beta$
decay. For detectors with higher density, such as discrete detectors
or liquid scintillator detectors, these electrons deposit their energy
within a few millimeters, allowing a less powerful but still useful
discrimination between ``compact'' signal-like events and
$\gamma$~rays, which are likely to scatter and deposit energy at
multiple sites. The difference may be resolved through discriminating
between the ``single-site'' and ''multi-site'' events by pulse-shape
discrimination or reconstructed event topology, depending on the
position resolution, as well as the size and type of a given
detector. Some detectors are capable of particle discrimination
through multiple detection channels, e.g. scintillation and ionization. 
This could allow for the identification of $\alpha$ backgrounds.

Timing is yet another key variable for distinguishing signal from
backgrounds. For example, in the aforementioned $^{222}$Rn chain, the
particularly troublesome $^{214}$Bi progeny decays in coincidence with
$^{214}$Po $\alpha$~decay, which has a 160-$\mu$s half-life. For some
detectors, this timing coincidence can be used to identify $^{214}$Bi
decays both in the bulk material and on the surfaces.

The spatial distribution of background events can be quite different
from that of the signal events. The $0\nu\beta\beta$-decay events will
be uniformly distributed throughout the source material, as will
background events from $2\nu\beta\beta$ decay and other uniformly
distributed radioactive sources. However, additional backgrounds will
come from the mechanical support materials and localized detector
components. Background events will be concentrated close to those
non-active materials.  In experiments with discrete detectors, each
detector may serve as a veto for other detectors in the system;
multiple-scattered background $\gamma$~rays or $\beta\gamma$ decays
are likely to deposit energy in more than one detector.  In monolithic
detectors, these background events may be rejected by an optimized
fiducial volume cut.  This configuration also allows the measurement of these
backgrounds with high statistics, which can in turn be used as a
constraint in the $0\nu\beta\beta$-decay analysis.

Many experiments use multiple variables to distinguish between signal
and backgrounds. While this can be accomplished with hard cuts or a
multi-dimensional fit, another option is to create an optimized
discriminator variable based on machine learning techniques. In future
searches, deep learning methods may also be applied to the problem of
signal-to-background optimization~\cite{Renner_2017}.

Another technique that can distinguish $0\nu\beta\beta$ decay from all
backgrounds other than $2\nu\beta\beta$ decay is the identification of
the decay daughter on an event-by-event basis. The prototypical
isotope for this technique is $^{136}$Xe~\cite{Moe:1991ik}. The
$\beta\beta$ decay of $^{136}$Xe results in an ionized Ba
daughter. This has been an intriguing system, with efforts by both the
nEXO~\cite{PhysRevA.76.023404,Chambers:2018srx} and
NEXT~\cite{PhysRevLett.120.132504} collaborations to identify single
Ba ions with high efficiency. This technique still presents
significant challenges to implementation, but the implication for 
positive identification of a $\beta\beta$ decay and the background 
rejection capabilities are significant enough to motivate continued
development for deployment in future experiments.

\section{\label{sec:DetTech}The Experimental Program}

Since the first direct searches for $0\nu\beta\beta$
decays~\cite{derMateosian:1966qz,Lazarenko:1966,Fiorini:1967in} in the
1960s, the experiments have grown from deploying grams to hundreds of
kilograms of decay isotopes.  As these detectors become more
sophisticated---from reducing the overall backgrounds to improving the
signal detection efficiency---the $T_{1/2}^{0\nu}$ limit has also
improved from ${\ls}10^{20}$~y to ${\gs}10^{26}$~y.

\begin{figure}[t]
\includegraphics[width=5.5in]{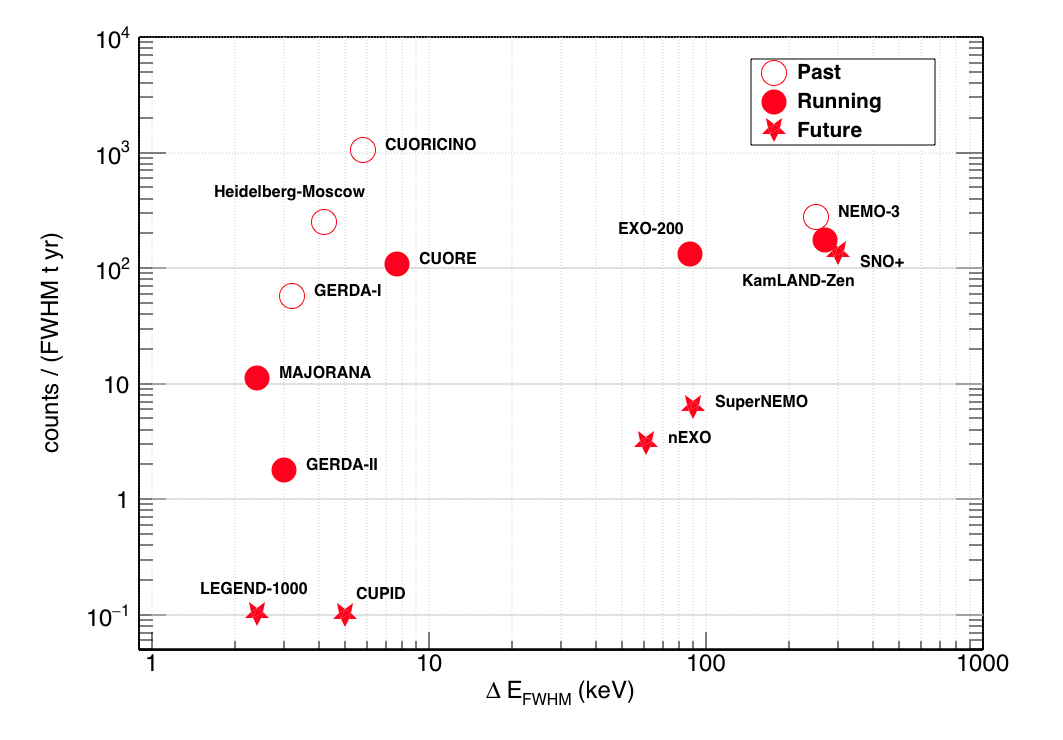}
\caption{The background index as a function of full-width-at-half-maximum energy resolution for selected past, current and future experiments with $^{76}$Ge, $^{100}$Mo, $^{130}$Te, and $^{136}$Xe as target. Note that large homogeneous detectors like SNO+, KamLAND-Zen, and nEXO are not well characterized by a single background index.}
\label{fig:expBgd}
\end{figure}

Much experimental progress has been made since the publication of the
last $0\nu\beta\beta$-decay review in this {\it Annual Review}
series~\cite{Elliott:2002xe}.  Table~\ref{tbl:curLimit} summarizes the
current lower limits in $T_{1/2}^{0\nu}$ and $\langle m_{\beta\beta}
\rangle$ for the different $\beta\beta$-decay isotopes.  It should be
kept in mind that there are many possible mechanisms for
$0\nu\beta\beta$ decay (Sec.\ \ref{sec:alt}); only the 
$\langle m_{\beta\beta} \rangle$ limits in the light-neutrino model
are summarized in the table.  Figure~\ref{fig:expBgd} shows two influential
detector parameters (Eq.~\ref{eq:Tsens}), energy resolution and
background index, for some of the past, current and future
experiments.  We have witnessed a tremendous amount of progress in
background reduction, but formidable challenges to improve further lie
ahead.  In the rest of this section, we will discuss the detector
technologies and the experimental program that are being pursued for
the discovery of $0\nu\beta\beta$-decay.

\begin{table}
\caption{\label{tbl:curLimit} $T_{1/2}^{0\nu}$ and $\langle m_{\beta\beta} \rangle$ limits (90\% C.L.) from the most recent 
measurements, sorted by the mass number. The $\langle m_{\beta\beta} \rangle$ limits are listed as reported in 
refereed publications.  Other unpublished preliminary results are described in the text. }
\begin{center} 
\begin{tabular}{llllc} \\ \hline\hline
Isotope           &  T$_{1/2}^{0\nu}$ ($\times 10^{25}$~y)                       & $\langle m_{\beta\beta}\rangle$ (eV)     &  Experiment &  Reference     \\ \hline
$^{48}$Ca       &  $>5.8\times 10^{-3}$                &  $<3.5-22$                           &   ELEGANT-IV &  \cite{Umehara:2008ru}             \\
$^{76}$Ge         &  $>8.0$                       &  $<0.12-0.26$                          &  GERDA &  \cite{Agostini:2018tnm} \\
                      &   $>1.9$                       &  $<0.24-0.52$                     &  {\sc Majorana Demonstrator} &  \cite{Aalseth:2017btx}  \\
$^{82}$Se        &  $>3.6\times 10^{-2}$                 &  $<0.89-2.43$                    &  NEMO-3 &  \cite{Barabash:2010bd}  \\
$^{96}$Zr	   & 	$>9.2 \times 10^{-4}$ &  $<7.2-19.5$                    &  NEMO-3 &  \cite{Argyriades:2009ph}  \\
$^{100}$Mo     &  $>1.1\times 10^{-1}$                       &  $<0.33-0.62$                           &  NEMO-3 &  \cite{Arnold:2015wpy}  \\
$^{116}$Cd       &  $>1.0\times 10^{-2}$                         &  $<1.4-2.5$                           &  NEMO-3 &  \cite{Arnold:2016bed}  \\
$^{128}$Te        &  $>1.1\times 10^{-2}$                       &  ---                     &  ---  &  \cite{Arnaboldi:2002te} \\
$^{130}$Te         &  $>1.5$                       &  $<0.11-0.52$                       &  CUORE &  \cite{Alduino:2017ehq}\\
$^{136}$Xe      &  $>10.7$                       &  $<0.061-0.165$                       &  KamLAND-Zen &  \cite{KamLAND-Zen:2016pfg} \\ 
    &  $>1.8$                       &  $<0.15-0.40 $                       &  EXO-200 &  \cite{Albert:2017owj} \\ 
$^{150}$Nd    &  $>2.0\times 10^{-3}$                       &  $<1.6-5.3$                             &  NEMO-3 &  \cite{Arnold:2016qyg} \\ \hline
\end{tabular}
\end{center}
\end{table}

\subsection{Semiconductors}

Among the different semiconductor detector technologies, 
$^{76}$Ge-enriched high-purity germanium (HPGe) detectors are
one of the most auspicious for scaling to a tonne-scale experiment.
The advent of using HPGe detectors in $\gamma$-ray spectroscopy and
the network of commercial manufacturers have propelled this technology
to a mature state.  Other semiconductor technologies, 
e.g.\ CdZnTe~\cite{Ebert:2015rda} or a recent idea of a complementary
metal-oxide-semiconductor (CMOS) pixel array~\cite{Chavarria:2016hxk},
are still in an early feasibility study stage and are unlikely to be
realized as a next-generation tonne-scale experiment.

There are several advantages of using HPGe detectors in
$0\nu\beta\beta$-decay searches. They are intrinsically clean as
impurities are removed in the detector crystal growing process.  They
can be fabricated with $^{76}$Ge-enriched materials; this
source-as-detector configuration enhances the signal detection
efficiency.  They have superior energy resolution; in fact, an energy
resolution of 0.12\%~(FWHM) at $Q_{\beta\beta}$ has been
attained~\cite{Aalseth:2017btx}.  However, HPGe detectors must be
fabricated and installed individually, which complicates the scaling
up to a larger array.

There have been a number of advances in HPGe detector design since the
last generation of $^{76}$Ge experiments,
Heidelberg-Moscow~\cite{KlapdorKleingrothaus:2000sn} and
IGEX~\cite{Aalseth:2002rf}.  The current generation of
experiments---GERDA~\cite{Ackermann:2012xja} and {\sc Majorana
  Demonstrator}~\cite{Abgrall:2013rze}---uses point-contact-type HPGe
detectors~\cite{Luke34577,Barbeau:2007qi} that have good
discrimination power between single-site signal and multi-site
background events.  These detectors also have very low capacitance,
allowing sensitive probe of new physics, such as dark matter searches,
at energies much lower than $Q_{\beta\beta}$~\cite{Abgrall:2016tnn}.

\subsubsection{$^{76}$Ge: GERDA}

The GERmanium Detector Array (GERDA)
experiment~\cite{Ackermann:2012xja} is located at the Laboratori
Nazionali del Gran Sasso (LNGS) in L’Aquila, Italy.  Bare 86\%-enriched
$^{76}$Ge HPGe detectors are immersed in a liquid-argon cryostat to
minimize the amount of nearby mechanical components and high-$Z$
shielding.  In its first phase, GERDA-I, 17.8~kg of the enriched
coaxial detectors from the Heidelberg-Moscow and IGEX experiments were
initially deployed, and were augmented by 3.63~kg in five ``Broad
Energy Ge'' (BEGe) p-type point-contact detectors about halfway
through data taking.  After accumulating an exposure of 21.6~kg~y of
data from late 2011 to mid-2013, the GERDA-I
results~\cite{Agostini:2013mzu} refuted the controversial claim of a
4.2$\sigma$ significance of observing $0\nu\beta\beta$ decay in
$^{76}$Ge~\cite{KlapdorKleingrothaus:2004wj}.  An
upgrade~\cite{Agostini:2017hit} prior to the second phase GERDA-II
improved the radioactivity backgrounds and their rejection, as well as
increasing the total enriched detector mass to 35.8~kg, out of which
BEGe detectors comprised 20~kg.  The background improvements include
the installation of a scintillating nylon shroud to shield the
detectors from $^{42}$K in the liquid argon; $^{42}$K is a progeny 
of $^{42}$Ar and has a maximum $\beta$ energy of
3525~keV~\cite{Lubashevskiy:2017lmf}.

An efficient active veto, coupled with effective pulse-shape
discrimination algorithms~\cite{Agostini:2013jta,Budjas:2009zu} to
reject multi-site and $\alpha$ background events enabled 
GERDA-II to achieve an unprecedentedly low background index of ($1.0^{+0.6}_{-0.4})\times 10^{-3}$~count/(keV~kg~y)~\cite{Agostini:2018tnm}.  Combining the
results from both phases, GERDA achieved a lower limit of
$T_{1/2}^{0\nu}$ at~$8.0\times 10^{25}$~y~(90\%~C.L.).  The median
sensitivity assuming null signal was $5.8\times 10^{25}$~y.

In mid-2018, further improvements to the scintillation-light detection
efficiency and the overall performance of the HPGe detectors were
implemented.  The collaboration installed five enriched ``inverted
coax point-contact'' (ICPC) detectors~\cite{Cooper:2011} with a total
mass of 9~kg; these ICPC detectors are a promising candidate for
future $^{76}$Ge $0\nu\beta\beta$-decay experiments.

\subsubsection{$^{76}$Ge: {\sc Majorana Demonstrator}}

The {\sc Majorana Demonstrator} (MJD) experiment~\cite{Abgrall:2013rze} 
is operating 29.7~kg of 88\%-enriched $^{76}$Ge and 14.4~kg of natural 
p-type point-contact detectors at the 4850-ft level of the Sanford Underground 
Research Facility~\cite{Lesko:2015sma} in Lead, SD, USA.

Unlike the GERDA experiment, the MJD experiment opted for a
traditional arrangement that the detectors are installed in two copper
vacuum cryostats, which are encapsulated in a graded shield consisting
of layers of copper, lead, an active muon veto, polyethylene and
borated polyethylene.  The whole setup is enclosed in a radon
exclusion box.  The experiment relied on using ultra-clean materials
and process control~\cite{Christofferson:2017nih} to reach the
background objectives.  For example, the copper used in the cryostat,
small parts near the detectors and the innermost copper shield were
electroformed~\cite{Hoppe:2014nva} and machined in the underground
cleanroom to prevent contamination and cosmogenic activation.

The MJD experiment has been taking data since 2015 when only the first
of two cryostats was populated with enriched detectors and the
construction of the graded shield was just beginning.  The
construction was fully complete by early 2017.  The collaboration has
released the $0\nu\beta\beta$-decay search results for these different
experimental configurations.  With an integrated exposure of
26~kg~y in the latest data release, the MJD collaboration obtained a lower
limit of $T_{1/2}^{0\nu}$ at~$2.7\times 10^{25}$~y~(90\%~C.L.) with the
median sensitivity of $4.8\times 10^{25}$~y~\cite{Alvis:2019sil}.  
The measured energy resolution is $2.53\pm 0.08$~keV (FWHM), and the
background index in the lowest-background experimental
configuration is ($4.7\pm 0.8) \times 10^{-3}$~count/(keV~kg~y). 
Similar to the GERDA analysis, pulse-shape discrimination techniques
to identify and reject multi-site events~\cite{Alvis:2019dzt} and
$\alpha$ backgrounds~\cite{Gruszko:2016hfh} were implemented.

\subsubsection{$^{76}$Ge: LEGEND}

The GERDA and MJD results have demonstrated the technical feasibility
to build a large-scale $^{76}$Ge-based $0\nu\beta\beta$-decay
experiment with ultra-low background and superior energy resolution.
The Large Enriched Germanium Experiment for Neutrinoless double-beta
Decay (LEGEND) collaboration was recently formed to pursue a
tonne-scale $^{76}$Ge-based experiment~\cite{Abgrall:2017syy}.  The
project will combine the strengths of the two operating
experiments---low-$Z$ shielding and scintillating veto for background
suppression from GERDA, and ultra-pure materials and components from
MJD---to attempt a $T_{1/2}^{0\nu}$ discovery sensitivity of
${\sim}10^{28}$~y in a phased program.

In the first phase, LEGEND-200, the GERDA experimental infrastructure
at LNGS will be modified and repurposed to accommodate up to
about 200~kg of $^{76}$Ge-enriched detectors.  The $T_{1/2}^{0\nu}$
discovery potential for LEGEND-200 is expected to be ${\sim}10^{27}$~y
with a background index of 0.6~count/(FWHM~t~y), a factor of ${\sim}$5
reduction from that in GERDA.  As of this writing, LEGEND-200 has been
nearly fully funded and the operation is anticipated to start in
2021. To reach the ultimate discovery potential at $T_{1/2}^{0\nu}
\sim 10^{28}$~y, the background index in the \numprint{1,000}~kg of
detectors in the subsequent phase, LEGEND-1000, needs to be further
reduced to ${\ls}$0.1~count/(FWHM~t~y).

\subsection{Bolometers}

The bolometric technique was first proposed for $0\nu\beta\beta$-decay
search in 1984~\cite{FIORINI198483}. Bolometers are cryogenic calorimeters
that operate at temperatures of ${\sim}10$~mK. An absorber is
connected to a low-temperature thermal bath via a weak thermal link,
and the temperature is read out by a sensitive thermometer.

Bolometer absorbers can be grown from a wide variety of materials,
including multiple $\beta\beta$-decay isotopes. Examples of these
crystals include TeO$_2$ (natural or enriched in $^{130}$Te),
$^{116}$CdWO$_4$, Zn$^{82}$Se, $^{40}$Ca$^{100}$MoO$_4$,
Zn$^{100}$MoO$_4$, and Li$_2^{100}$MoO$_4$.  Like semiconductor
detectors, crystalline bolometers can be intrinsically low in
radioactivity because of the crystal growth process. The readers are
referred to a comprehensive review of the use of bolometers in
$\beta\beta$-decay experiments in Ref.~\cite{Poda:2017jnl}.

The typical rise in temperature is of the order of ${\sim}0.1$~mK
per~MeV of deposited energy. Highly sensitive thermometers such as
neutron-transmutation-doped (NTD) germanium or silicon, transition
edge sensors (TES), metallic magnetic calorimeters (MMC), and kinetic
inductance detectors (KID) are used for reading out such minuscule
temperature changes. The NTD~Ge thermistors are the most widely 
used in $0\nu\beta\beta$-decay searches. In future
bolometer-based experiments, the TES and KID devices will become more
important as they can be used to detect the Cherenkov or scintillation
light due to radioactive backgrounds in the crystals; thus, allowing
the associated event to be rejected.
 
Excellent counting statistics in the phonon channel imply that
bolometers should have comparable energy resolution to semiconductor
detectors. They are inherently segmented arrays of crystals, like the
semiconductor detectors, and therefore do not benefit dramatically
from self shielding as detector size increases. The challenge of
working at extreme low temperature increases the technical difficulty
of building large detectors.

\subsubsection{$^{130}$Te: CUORE}

The Cryogenic Underground Observatory for Rare Events (CUORE)
experiment is located at LNGS.  It consists of a close-packed array of
988 $5{\times}5{\times}5$~cm$^3$, 750-g TeO$_2$ absorber crystals
arranged into 19 towers and cooled to 7~mK by a powerful dilution
refrigerator. Like the MiDBD and Cuoricino~\cite{Arnaboldi:2004qy},
and CUORE-0~\cite{Alduino:2016vjd} experiments that preceded it, the
CUORE experiment uses unenriched Te, taking advantage of the large
natural abundance of $^{130}$Te~\cite{ARNABOLDI2004775}. The
absorbers are instrumented with NTD~Ge thermistors that are read out
continuously. Each crystal is also outfitted with a heater for thermal 
gain stabilization, and further calibration is provided by $\gamma$-ray 
sources deployed between the towers. The array is surrounded by layers of 
$\gamma$-ray and neutron shielding, including low-background ancient 
Roman lead shields in the cryogenic volume. Additional lead, 
borated polyethylene, and boric acid are located outside the cryostat 
for additional shielding.

The first $0\nu\beta\beta$-decay search results from the CUORE 
experiment, based on two month-long runs for a total exposure 
of 24.0~kg~y of $^{130}$Te, set a limit of
$T_{1/2}^{0\nu}>1.3\times10^{25}$~y~(90\%~C.L.) with
the median sensitivity of $7.0\times10^{24}$~y~\cite{Alduino:2017ehq}. 
In combination with the previous results from Cuoricino and CUORE-0, 
the limit becomes $T_{1/2}^{0\nu}>1.5\times10^{25}$~y~(90\%~C.L.). 
The experiment has achieved an energy resolution of 
$7.7\pm0.5$~keV~(FWHM), or 0.30\%, at $Q_{\beta\beta}$ and a background of
$0.014\pm0.002$~count/(keV~kg~y). The projected sensitivity of the
CUORE experiment is $9\times10^{25}$~y after five years of
running~\cite{Alduino2017}.

\subsubsection{$^{82}$Se/$^{100}$Mo/$^{130}$Te: CUPID}

While the CUORE detector reads out a single energy signal and
therefore has minimal background discrimination capabilities, the
CUORE Upgrade with Particle IDentification (CUPID) collaboration is
exploring bolometer development to improve background rejection
through active particle discrimination~\cite{ARTUSA2017321},
particularly the rejection of the dominant $\alpha$ backgrounds in
CUORE~\cite{Wang:2015raa}. One approach is to detect the small
Cherenkov-light signal in TeO$_2$~\cite{Battistelli:2015vha}.  The
use of $^{130}$Te-enriched bolometers would further extend 
the reach of this CUPID configuration.

Another approach is to deploy scintillating bolometers, such as the
Zn$^{82}$Se crystals in CUPID-0~\cite{PhysRevLett.120.232502}, or the
Zn$^{100}$MoO$_4$ and Li$_2^{100}$MoO$_4$ crystals in the LUMINEU
experiment~\cite{Barabash:2014una,Poda:2017bmd}.  CUPID-Mo, an
experiment evolved from LUMINEU, has been running twenty
$^{100}$Mo-enriched 0.2-kg Li$_2^{100}$MoO$_4$ crystals in the
cryogenic setup of the EDELWEISS dark matter experiment at Laboratoire
Souterrain de Modane (LSM) in the Fr\'{e}jus Tunnel near Modane,
France~\cite{Poda:2017bmd}.  Additional detectors will be deployed in
the CUPID-0 setup at LNGS in 2019.  The outcome of these research and
development efforts will decide the best technology for the
tonne-scale CUPID program that has the goals of a background index of
${\sim}0.1$~count/ROI~t~y and 
$T_{1/2}^{0\nu} > 10^{27}$~y~\cite{ouellet_jonathan_2018_1286904}.

\subsubsection{$^{100}$Mo: AMoRE}

The Advanced Molybdenum based Rare process Experiment (AMoRE) is a
$^{100}$Mo-based experiment at the Yangyang Underground Laboratory
(Y2L) in South Korea.  It comprises calcium molybdate scintillating
crystals that are depleted to ${\sim}0.002$\% in $^{48}$Ca but
enriched to ${\sim}95$\% in $^{100}$Mo~\cite{Lee:2018gcu}.  Metallic
magnetic calorimeter sensors are used to read out the phonon signals.
One of the two MMC sensors on each crystal is coupled to a gold film
on a germanium wafer.  The phonons generated from the light absorbed
in the wafer are collected by the gold film and measured by the
attached sensor via a superconducting quantum interference device
(SQUID).  AMoRE-Pilot, the pilot phase of the project, has been
operating since 2015. AMORE-I and AMORE-II, the next phases of the
project with ${\sim}$5~kg of $^{48\mathrm{depl.}}$Ca$^{100}$MoO$_4$
crystals and ${\sim}$200~kg of $^{100}$Mo-based crystals, are
projected to reach a $T_{1/2}^{0\nu}$ sensitivity of ${\sim}10^{25}$~y
and ${\sim}5 \times 10^{26}$~y,
respectively~\cite{kim_seungcheon_2018_1300715}.

\subsection{Time Projection Chambers}

The time projection chamber (TPC) is an attractive detector technology
for $0\nu\beta\beta$-decay searches because of a combination of mass
scalability and access to multiple background discrimination
variables. A TPC takes advantage of a detection medium that produces
two energy channels: ionization and scintillation. The combination of
these two signals allows the reconstruction of event topology,
position, and energy. The ionization-to-scintillation ratio provides
convenient particle discrimination between $\alpha$~particles (high
recombination leading to low ionization to scintillation) and
$\gamma$~rays or $\beta$~electrons (low recombination leading to
relatively high ionization to scintillation). For
$0\nu\beta\beta$-decay searches, $^{136}$Xe-enriched xenon is a
convenient source and detection medium. Xenon TPCs can be built for
both gas and liquid phases.  By operating high-pressure gas-phase
xenon TPCs in electroluminescent mode, an energy resolution of better
than 0.5\% FWHM at $Q_{\beta\beta}$ can be
achieved~\cite{Alvarez_2012}.

Liquid-phase xenon TPCs offer maximum source density. Two-phase
liquid-xenon detectors are popular for dark matter searches. These
experiments---LUX-ZEPLIN~\cite{Akerib:2015cja} and
XENON-nT~\cite{2017APS..APR.J9003A} that are under construction, and
the future DARWIN~\cite{Aalbers_2016} project---might have the
capability to search for $0\nu\beta\beta$ decay.  In fact, DARWIN aims
at a $T_{1/2}^{0\nu}$ sensitivity of $8.5 \times 10^{27}$~y (90\%
C.L.) for a natural xenon exposure of 140~t~y, which is comparable to
dedicated tonne-scale $0\nu\beta\beta$-decay experiments.

Rather than optimizing for a low energy threshold as in dark matter
detectors, liquid-phase xenon TPCs for $0\nu\beta\beta$-decay searches
are optimized for low-radioactive-background construction and energy
resolution, resulting in the choice of single-phase detectors. The
achievable energy resolution is somewhat worse than that of the
gas-phase detectors.  While scattering prevents the resolution of the
two $\beta$ tracks, multi-site background and spatial distribution
discrimination work well with position resolution achievable at the
few-mm level.

\subsubsection{$^{136}$Xe: EXO-200}

EXO-200~\cite{Auger_2012}, a prototype of the Enriched Xenon
Observatory (EXO) project, was located at the Waste Isolation Pilot
Plant near Carlsbad, NM, USA. The cylindrical single-phase liquid-xenon
TPC was filled with an active mass of 110~kg of xenon, enriched to
80.6\% in $^{136}$Xe, at a temperature of
167~K~\cite{Auger_2012}. EXO-200 employed a central cathode with
detector planes at both ends consisting of crossed-wire grids for
ionization collection and large-area avalanche photodiodes for
scintillation collection. The low-mass copper vessel for xenon
containment was surrounded by HFE-700 cooling and shielding fluid 
within a double-walled copper cryostat, which was in turn inside 
25~cm of low-background lead shielding with an active
muon veto. An extensive screening program was undertaken to select
detector materials~\cite{Leonard:2007uv,Leonard:2017okt}. In addition,
EXO-200 analysis employs a multi-dimensional approach to background
discrimination, including spatial and topological information, as well
as particle discrimination.

EXO-200 data taking proceeded in two phases. Phase I began
taking data with enriched xenon in 2011 and reported the first
observation of $2\nu\beta\beta$ decay in $^{136}$Xe
\cite{PhysRevLett.107.212501}.  EXO-200 has produced a precision
measurement of the $2\nu\beta\beta$-decay half-life, demonstrating the
power of the liquid-xenon TPC technique~\cite{PhysRevC.89.015502}.
Phase I ended because of an unrelated fire and radiation release at
the experimental site in early 2014. The experiment was upgraded with
a radon suppression system and low-noise electronics. Data taking
restarted for Phase II in 2016 and completed in 2018. The first
results from Phase II gave a $T_{1/2}^{0\nu}$ limit of
$1.8\times10^{25}$~y (90\% C.L.)~\cite{Albert:2017owj}.  The current
EXO-200 detector performance displays energy resolution of
2.90\%~(FWHM) at $Q_{\beta\beta}$ and a background index of
$(1.6\pm0.2)\times10^{-3}$~/(keV~kg~y) in the $\pm 2\sigma$ ROI. Final
analysis of the full EXO-200 dataset is in progress.

\subsubsection{$^{136}$Xe: nEXO}

nEXO is a planned tonne-scale single-phase liquid-xenon TPC based on
the success of EXO-200~\cite{Kharusi:2018eqi}. The TPC will contain
5000~kg of xenon enriched to 90\% in $^{136}$Xe. With lower noise
silicon photomultipliers (SiPMs) for scintillation collection, the
expected energy resolution will be 2.4\%~(FWHM) at
$Q_{\beta\beta}$. Multiple underground locations for hosting the nEXO
experiment have been studied, including the SNOLAB
Cryopit~\cite{Duncan:2010zz}. The projected $T_{1/2}^{0\nu}$
sensitivity for the experiment is approximately $10^{28}$~y with a
$3\sigma$ discovery potential of
$5.7\times10^{27}$~y~\cite{PhysRevC.97.065503}. At this sensitivity
and with the projected energy resolution, the $2\nu\beta\beta$-decay
background is negligible. The power of this detector comes from having
a large monolithic source volume, good energy resolution, and the
background discrimination capabilities of a TPC. The low-background
materials and construction techniques needed to achieve this
sensitivity are not beyond what has already been demonstrated by
current experiments.

\subsubsection{$^{136}$Xe: NEXT}

The Neutrino Experiment with a Xenon TPC (NEXT) is a planned
high-pressure gas-phase xenon TPC that employs amplification via
electroluminescence to achieve an energy resolution of ${<}$1\% (FWHM)
at $Q_{\beta\beta}$~\cite{Mart-Albo2016}. NEXT-100, which will deploy
100~kg of enriched xenon at 15~bar, will be located at the Laboratorio
Subterr\'{a}neo de Canfranc (LSC) in Spain.  At this pressure,
individual $\beta$ tracks can be resolved, including increased energy
deposition at the end of the track where the electron becomes
non-relativistic. This distinctive topological signature for
$\beta\beta$-decay events can be used to reject other sources of
background.

An array of photomultiplier tubes (PMTs) detects both the primary
scintillation light and the secondary scintillation light to
reconstruct the event energy, and a second array of SiPMs
located near the amplification region is used for track
reconstruction. Tracking allows background rejection through the
identification of individual $\beta$ energy depositions in the
detector.  Initial studies estimate that a signal efficiency of 28\%
and a background rate of $4\times10^{-4}$~count/(keV~kg~y) are
achievable. The NEXT-100 detector is projected to reach a
$T_{1/2}^{0\nu}$ sensitivity of $2.8\times10^{25}$~y (90\%~C.L.) after
three years of running.

\subsubsection{$^{136}$Xe: PandaX-III}

The Particle and Astrophysical Xenon Experiment III (PandaX-III),
located at the China Jinping Underground Laboratory II (CJPL-II), is a
high-pressure gas-phase TPC for $0\nu\beta\beta$-decay search in
$^{136}$Xe~\cite{Han:2017fol}.  Its first phase will feature one
200-kg TPC module operated at a pressure of 10~bar.  The next phase
will comprise five upgraded modules, bringing the experiment to tonne
scale.  Charges are read out by microbulk micromegas (MM)
modules~\cite{Andriamonje:2010zz} that line the end caps of the
cylindrical vessel.  The expected energy resolution and background
index for the 200-kg module are 3\%~(FWHM) at $Q_{\beta\beta}$ and
${\sim}10^{-4}$~count/(keV~kg~y) in the ROI, respectively.  The
projected $T_{1/2}^{0\nu}$ sensitivity is $10^{26}$~y after three
years of running.  With an improved energy resolution of 1\%~(FWHM)
and a lower background index of ${\sim}10^{-5}$~count/(keV~kg~y), the
tonne-scale PandaX-III would reach a $T_{1/2}^{0\nu}$ sensitivity of
$10^{27}$~y after three years.

\subsection{Organic Scintillators}

Although organic scintillators do not have superior energy resolution,
their main appeal as a $0\nu\beta\beta$-decay detector is the mass
scalability.  Unlike other $0\nu\beta\beta$-decay experiments with 
solid targets, contaminants in the liquid scintillator may be
removed online.  The $\beta\beta$-decay isotope can be removed during
circulation as well, allowing possible systematic checks of rate
scaling in the event of a discovery.

There are typically two components in liquid scintillators: solvents
that form the bulk, and fluors with an emission spectrum that better
matches the response of the photodetectors as a dopant.  The popular
choices of solvents in previous large-scale neutrino experiments were
pseudocumene and dodecane, e.g.~KamLAND used a 20:80 by volume mix of
the two solvents.  A solvent that is gaining popularity in recent
years is linear alkylbenzene~(LAB)~\cite{OKeeffe:2011dex}.  This
solvent has a high flash point, low toxicity, high compatibility with
most materials, and low cost.  A common choice for fluor in
$\beta\beta$-decay experiments is PPO (2,5-diphenyloxazole).  Both
KamLAND-Zen ($^{136}$Xe) and SNO+ ($^{130}$Te) experiments use this
wavelength-shifting agent.

As we discussed in Sec.~\ref{sec:expbgd}, the irreducible background
of solar neutrinos scattering off atomic electrons is problematic for
large liquid scintillator detectors. These elastic scattering events
have strong directionality and are correlated to the Sun's direction.
The key to mitigate this background is to 
separate the directional Cherenkov light emitted by the relativistic
electrons from the isotropic scintillation light that has a much
higher intensity.  In bench measurements, the
CHESS~\cite{Caravaca:2016ryf} and FlatDot~\cite{Gruszko:2018gzr}
experiments have recently demonstrated the separability of the prompt
Cherenkov and the delayed scintillation light through timing in
LAB-based scintillators.  These encouraging results could lead to the
realization of much larger detectors for $0\nu\beta\beta$-decay
searches, such as the proposed 50-kilotonne THEIA
detector~\cite{Gann:2015fba}.

\subsubsection{$^{136}$Xe: KamLAND-Zen and KamLAND2-Zen}

Using the KamLAND (``Kamioka Liquid scintillator AntiNeutrino
Detector'') infrastructure at the Kamioka Observatory in the Gifu
prefecture in Japan, the KamLAND-Zen (''Zero neutrino'') experiment
searches for $0\nu\beta\beta$ decay in $^{136}$Xe.  Various amounts of
$^{136}$Xe, enriched to 90\%, were at different times loaded in a
liquid scintillator cocktail of 82\% decane and 18\% pseudocumene
by volume, along with 2.7~g/l of PPO as 
fluor.  This xenon-loaded liquid scintillator (Xe-LS) is contained in
a 25-$\mu$m-thick nylon ``mini-balloon,'' suspended in liquid
scintillator at the center of the 13-m-diameter main balloon.  The
main balloon is installed in a 18-m-diameter stainless steel spherical
vessel, which is filled with a non-scintillating buffer oil.  On the
vessel, \numprint{1,879} 17-inch and 20-inch PMTs, combined to a photocathode
coverage of 34\%, are mounted.  A 3.2-kt cylindrical water-Cherenkov
detector outside the containment vessel serves as a muon veto.

In phase I of KamLAND-Zen~400, 320~kg of enriched xenon was loaded.  A
dominant background from $^{110\mbox{\tiny m}}$Ag $\beta$~decay, believed
to be the fallout from the Fukushima incident in 2011, limited the
sensitivity of the experiment~\cite{Gando:2012zm}.  The Xe-LS was
subsequently purified over 1.5 years and the contamination was reduced
by an order of magnitude successfully prior to the commencement of
phase II.  Various event selection criteria, including Bi-Po time
coincidence and a fudicial volume limited to the radial distance of
2~m from the center of the detector, were used to improve the
signal-to-noise ratio in the signal region of interest.  With 380~kg
of enriched xenon and a total exposure of 504~kg~y in phase II, the
KamLAND-Zen~400 experiment obtained a $T_{1/2}^{0\nu}$ lower limit of
$1.07\times 10^{26}$~y~(90\%~C.L.) and a median sensitivity of
$5.6\times 10^{25}$~y~\cite{KamLAND-Zen:2016pfg}.

Detailed background studies in phase II identified contaminations on
the surface of the mini-balloon, as well as residual
$^{110\mbox{\tiny m}}$Ag in the liquid scintillator.  An arduous effort to
purify the liquid scintillator and to remake the balloon ensued at the
end of phase-II running.  The collaboration has completed the
installation of a new mini-balloon, and will load 750~kg of enriched xenon in
this new phase of KamLAND-Zen~800 experiment imminently.

In the longer term, the KamLAND-Zen collaboration plans to deploy over
a tonne of enriched xenon and to reduce the $2\nu\beta\beta$-decay
background in the signal region of interest by improving the 
detector resolution in the KamLAND2-Zen experiment.  The 
research and development of various strategies to increase 
the amount of detected scintillation light---from increasing
the light yield with a different liquid scintillator, to increasing
the light collection with light concentrators and improving the
detection efficiency with PMTs that have higher quantum
efficiency---are being conducted.  Background reduction and rejection
studies, involving the development of new techniques in $^{10}$C
rejection and the fabrication of a scintillating mini-balloon, are in
progress as well.  If the goal to improve the energy resolution by a
factor of two were reached, KamLAND2-Zen may reach a $T_{1/2}^{0\nu}$
sensitivity of ${\sim}2\times 10^{27}$~y after five years of running.

\subsubsection{$^{96}$Zr: ZICOS} 

The Zirconium COmplex in liquid Scintillator (ZICOS)
experiment~\cite{1742-6596-718-6-062019} is a new effort to dissolve a
high concentration of tetrakis (isopropyl acetoacetato) zirconium
(Zr(iprac)$_4$) in liquid scintillator.  The collaboration is
investigating the properties of the liquid scintillator, including the
ability to separate the Cherenkov and scintillation light from 
$^{208}$Tl $\beta-\gamma$ decay.  Preliminary design studies indicate
that ${\sim}$45~kg of $^{96}$Zr enriched to 50\% can potentially reach a
$T_{1/2}^{0\nu}$ lower limit of $2\times 10^{26}$~y; the ability to
reject $^{208}$Tl $\beta-\gamma$ background in situ would improve 
the limit to ${\sim}10^{27}$~y.

\subsection{Inorganic Scintillators}

With a high $Q_{\beta\beta}$ of 4.27~MeV, inorganic CaF$_2$
scintillators have been receiving interests from experimenters since
the early days of $0\nu\beta\beta$-decay searches. The
$\beta\beta$-decay isotope $^{48}$Ca is amalgamated in the
scintillator crystal growing process.  The primary background in a
$^{48}$Ca experiment is no longer the 2615-keV $\gamma$~ray from
$^{208}$Tl; instead, the more penetrating $\gamma$ rays from
$(n,\gamma)$ radiative capture in the containment vessel and the rock
surrounding the crystals are the most
significant~\cite{Nakajima:2018okw}.  These high-energy $\gamma$ rays
can either be shielded by passive shielding or identified by a
an active veto, or both.  The most challenging aspect of a $^{48}$Ca
experiment is to find a cost effective process to enrich the isotope,
whose natural abundance is only 0.187\%.  A recent table-top
experiment has demonstrated a significant enrichment ratio using
multi-channel counter-current electrophoresis~\cite{KishimotoDBD18};
this breakthrough has the promise to produce significant quantities of
$^{48}$Ca at a much lower cost than other traditional enrichment
techniques.

\subsubsection{$^{48}$Ca: CANDLES}

The CANDLES series~\cite{Umehara:2015lla} of $0\nu\beta\beta$-decay
searches in $^{48}$Ca is being carried out at the Kamioka Observatory.
In the latest CANDLES-III setup, 96 natural CaF$_2$ scintillator
crystals with a total mass of 305~kg are suspended from the roof of a
2~m$^3$ liquid scintillator vessel.  Scintillation light from both the
inorganic crystals and the liquid scintillator is observed by 62~PMTs
via light pipes.  The PMTs are mounted in a 3-m~diameter, 4-m~tall
cylindrical water tank.  The experiment recently reported preliminary
results from 131~live~days of data~\cite{KishimotoDBD18}.
Contamination from Th up to ${\sim}$60~$\mu$Bq/kg was observed in some
of the crystals.  For those crystals with a Th contamination of
${<}$10~$\mu$Bq/kg, there is no candidate event in the signal region of
interest, resulting in a $T_{1/2}^{0\nu}$ lower limit of $6.2\times
10^{22}$~y~(90\%~C.L.) and a median sensitivity of $3.6\times
10^{22}$~y.

\subsection{Tracking Calorimeters} 

Tracking calorimeters take a multi-layered detection approach. Rather
than distributing the source throughout the detector volume, tracking
calorimeters like NEMO-3~\cite{ARNOLD200579} and
SuperNEMO~\cite{Arnold:2010tu} use a thin foil of source material in
the center of a sandwich configuration, surrounded first by a
low-pressure gas tracking layer to track the two $\beta$ particles and
then a calorimetric layer to measure the energy. This type of
detectors provides superior topological information and is the only
detector technology capable of measuring the opening angle between the
two $\beta$s---one observable that can distinguish certain underlying
mechanisms for $0\nu\beta\beta$ decay (Sec.\ \ref{sec:alt}). In
addition, many different isotopes can be formed into foils and studied
in the same detector configuration. Background discrimination is
excellent, but the thin source foils are difficult to scale up to
large exposure.

\subsubsection{$^{82}$Se and others: SuperNEMO}

SuperNEMO~\cite{Barabash:2017sxf,PatrickDBD18} is a next-generation
detector based on the technology demonstrated by
NEMO-3~\cite{ARNOLD200579}, which successfully studied multiple
$0\nu\beta\beta$ isotopes including $^{100}$Mo. The NEMO program is
unique in that the $\beta\beta$-decay source material is distinct from
the detection medium, allowing multiple isotopes to be studied with a
single detector configuration.  A demonstrator module for SuperNEMO is
under construction at the Laboratoire Souterrain de Modane. This
module contains 6.3~kg of $^{82}$Se in 34~foils, surrounded by a
tracking detector made up of drift cells operating in Geiger
mode. Operating in a magnetic field of 25~G, the tracking detector
allows for the identification of the two $\beta$ particles for both
background rejection and the measurement of angular correlations. The
tracking detector is surrounded on four sides by calorimetry planes
consisting of blocks of scintillator read out by PMTs. The planned
energy resolution of the detector is 4\%~(FWHM) at 3~MeV, the
$Q_{\beta\beta}$ of $^{82}$Se. The demonstrator module will reach a
$T_{1/2}^{0\nu}$ sensitivity of ${>}5.85\times 10^{24}$~y (90\% C.L.)
after 2.5~years of running.  For a full SuperNEMO detector consisting
of 20 modules, the sensitivity to the half-life of $^{82}$Se is
projected to be $1.2\times10^{26}$~y.

\section{\label{sec:conclusions}Conclusions}

We have described the most recent theoretical and experimental
development in $0\nu\beta\beta$ decay.  If discovered, this
lepton-number-violating process would have profound implications for
our understanding of the evolution of the Universe and the fundamental
theory of elementary particles.

In the light-neutrino exchange mechanism, neutrino mass limits
approaching the projected mass-scale sensitivity of the KATRIN
experiment are around the corner. Tests of the inverted-ordering
regime are crucial for the standard neutrino paradigm, and those in
the range of normal ordering are even more so. Beyond the
light-neutrino regime, there are various plausible mechanisms spanning
a multitude of energy scales, including those accessible 
at present and future colliders, that could mediate $0\nu\beta\beta$
decay.  This implies interesting tests of the mechanisms and
investigations of the ``inverse problem'' of the decay,
i.e.\ identifying the origin of the decay once it is observed.  The
nuclear and hadronic aspects of $0\nu\beta\beta$ decay remain the
complications for precise physics extraction from an experimental limit or
a possible signal. In the development of new theoretical approaches,
new aspects of these problems are being routinely discovered. These
are signs of a vibrant field, and give hope that those uncertainties
will become much smaller in the near future.

On the experimental side, this is an exciting time to search for
$0\nu\beta\beta$ decay, the only realistic direct probe for
lepton-number violation.  Since the first direct searches in the
1960s, the $T_{1/2}^{0\nu}$ limit has improved by six orders of
magnitude, reaching ${\gs}10^{26}$~y in the current generation of
experiments.  These experiments feature the deployment of different
$\beta\beta$-decay isotopes and detector technologies, enabling
numerous advancements in isotope preparation, clean-material
development, radioactivity mitigation, signal detection, and
analysis.

Well-motivated half-life predictions are within experimental reach.
The next-generation of $0\nu\beta\beta$-decay searches has the
potential of a discovery at $T_{1/2}^{0\nu}$ exceeding $10^{28}$~y.
To realize this goal, these experiments will have to be able to
achieve the formidable background index of ${\ls}0.1$~count/(FWHM~t~y)
in a robust tonne-scale detector that is expected to operate with a
high duty cycle for a decade or longer.  Despite these challenges,
international teams are spearheading efforts to mount at least one of
these experiments projecting to reach the intermediate
$T_{1/2}^{0\nu}$ discovery potential of ${\sim}10^{27}$~y in the
coming decade.

\section*{DISCLOSURE STATEMENT}
The authors are not aware of any affiliations, memberships, funding,
or financial holdings that might be perceived as affecting the
objectivity of this review.

\section*{ACKNOWLEDGMENTS}
We thank Brian Fujikawa, Wick Haxton, Martin Hirsch, and Tadafumi
Kishimoto for helpful discussions. We are grateful for Ralph
Massarczyk's help in producing Fig.~\ref{fig:expBgd}.
MD is partially supported by the U.S. Department of Energy (DOE), Office of
Science, Office of Nuclear Physics under Award No.\ DE-SC0014517 to
Drexel University.  WR is supported by the DFG with grant RO 2516/7-1
in the Heisenberg Programme. Lawrence Berkeley National Laboratory
(LBNL) is operated by The Regents of the University of California (UC)
for the U.S.\ DOE under Federal Prime Agreement DE-AC02-05CH11231.

\bibliography{refs}

\end{document}